\documentclass[twocolumn]{aastex62}

\usepackage[utf8]{inputenc}
\usepackage{newunicodechar}
\usepackage{booktabs}
\usepackage{multirow}
\usepackage{longtable}
\usepackage{array}
\usepackage{amsmath}

\newcommand{\textprime}{\ensuremath{'}}
\newunicodechar{′}{\textprime}

\shortauthors{Gootkin et al.}
\shorttitle{13 Years of P Cygni Spectropolarimetry}

\begin{document}

\title{13 Years of P Cygni Spectropolarimetry: Investigating Mass-loss Through H$\alpha$, Periodicity, and Ellipticity}

\author[0000-0003-0922-138X]{Keyan Gootkin}
\affiliation{Department of Astronomy, University of Washington, Seattle, WA, USA}
\author[0000-0003-3601-3180]{Trevor Dorn-Wallenstein}
\affiliation{Department of Astronomy, University of Washington, Seattle, WA, USA}
\author[0000-0001-8470-0853]{Jamie R. Lomax}
\affiliation{Physics Department, United States Naval Academy, 572C Holloway Rd, Annapolis, MD, 21402, USA}
\author[0000-0003-3734-8177]{Gwendolyn Eadie}
\affiliation{Department of Astronomy \& Astrophysics, University of Toronto, Toronto, ON, Canada}
\affiliation{Department of Statistical Sciences, University of Toronto, Toronto, ON, Canada}
\author[0000-0003-2184-1581]{Emily M. Levesque}
\affiliation{Department of Astronomy, University of Washington, Seattle, WA, USA}
\author{Brian Babler}
\affiliation{Department of Astronomy, University of Wisconsin-Madison, 475 N. Charter St., Madison, WI 53706, USA}
\author[0000-0003-1495-2275]{Jennifer L. Hoffman}
\affiliation{Department of Physics and Astronomy, University of Denver, 2112 East Wesley Ave., Denver, CO 80208, USA}
\author{Marilyn R. Meade}
\affiliation{Department of Astronomy, University of Wisconsin-Madison, 475 N. Charter St., Madison, WI 53706, USA}
\author{Kenneth Nordsieck}
\affiliation{Department of Astronomy, University of Wisconsin-Madison, 475 N. Charter St., Madison, WI 53706, USA}
\author{John P. Wisniewski}
\affiliation{Homer L. Dodge Department of Physics and Astronomy, University of Oklahoma, 440 W Brooks Street, Norman, OK 73019, USA}

\correspondingauthor{Keyan Gootkin}
\email{goot1024@uw.edu}
\begin{abstract}
    We report on over 13 years of optical and near-ultraviolet spectropolarimetric observations of the famous Luminous Blue Variable (LBV), P Cygni. LBVs are a critical transitional phase in the lives of the most massive stars, and achieve the largest mass-loss rates of any group of stars. Using spectropolarimetry, we are able to learn about the geometry of the near circumstellar environment surrounding P Cygni and gain insights into LBV mass-loss. Using data from the HPOL and WUPPE spectropolarimeters, we estimate the interstellar polarization contribution to P Cygni's spectropolarimetric signal, analyze the variability of the polarization across the H$\alpha$ emission line, search for periodic signals in the data, and introduce a statistical method to search for preferred position angles in deviations from spherical symmetry which is novel to astronomy. Our data are consistent with previous findings, showing free-electron scattering off of clumps uniformly distributed around the star. This is complicated, however, by structure in the percent-polarization of the H$\alpha$ line and a series of previously undetected periodicities. 
\end{abstract}

\section{Introduction}
Luminous Blue Variables (LBVs)---also called S Doradus or S Dor variables---are incredibly rare, with less than 20 confirmed members of this class in the Milky Way \citep{Richardson2018TheGroup}. Despite the short lifetime of the LBV phase ($\sim 10^5 yr$), these stars are an important set of objects for understanding the post-main sequence evolution of stars with initial masses $\geq 20 M_\sun$ \citep{Groh2014TheSpectra}. This eclectic group represents a critical transitional phase in the lives of the most massive stars.

LBVs reside at the top of the Hertzsprung-Russell diagram. This unique position---near the Humphreys-Davidson limit \citep{Humphreys1979STUDIESCLOUD}, and on the S Dor instability strip---produces strange behavior in these stars. LBVs are highly unstable, losing large amounts of mass, and dramatically varying both photometrically and spectroscopically. In their quiescent, hot state (12,000 to 30,000 K), LBVs resemble blue supergiants. However, during an outburst they move horizontally across the HR diagram to temperatures between 7500 K and 9000 K---maintaining a consistent bolometric luminosity even while $V$-band magnitudes fluctuate \citep{Humphreys1994TheGeysers}. 

Other than Wolf-Rayet stars, LBVs achieve the highest mass-loss rates of any group of stars, $10^{-5}-10^{-4}M_\sun$ \space $yr^{-1}$, and beyond the ``typical'' S Dor outbursts they undergo catastrophic (but non-terminal) eruptions in which significant mass is lost \citep{Humphreys1994TheGeysers}. The only observed eruptions in our galaxy were from P Cygni (also known as P Cyg, 34 Cyg, or Nova Cyg 1600), in which 0.1 $M_\sun$ was lost \citep{Smith2006InfraredOutburst}, and $\eta$ Car which is estimated to have lost over 10$M_\sun$ of material in its massive 19$^{th}$ century eruptions \citep{Smith2003MASSCARINAE}. However, the physical mechanisms behind the eruption, outburst, and quiescent modes of mass loss are poorly understood.

The subject of this paper, P Cygni, was discovered on August $18^{th}$, 1600. The Dutch cartographer, globe-maker, and former student of Tycho Brahe, Willem Janszoon Blaeu, observed a \textit{Nova Stella} in the heart of Cygnus \citep{Blaeu1602Himmelsglob}. He chronicled this discovery in an inscription on a celestial globe, made in his Amsterdam workshop in 1602\footnote{The first edition of this globe, as of the writing of this paper, is in the collections of Skokloster Castle, a castle and museum north of Stockholm and can be viewed online \href{http://emuseumplus.lsh.se/eMuseumPlus?service=ExternalInterface&module=collection&objectId=31659&viewType=detailView}{here}.}. This new star was P Cygni. Although it was called a nova at the time, it was later recognized as a Luminous Blue Variable (LBV).

During the 1600 eruption (the \textit{Nova Stella}), P Cygni brightened to the point of visibility (up to $3^{rd}$ magnitude) for the first time, and remained visible for 26 years before fading to $\sim6^{th}$ magnitude. It reached $\sim 3^{rd}$ magnitude again in 1654, experienced variability for a number of decades, but has been in a stage of quiescence since the late $18^{th}$ century \citep{deGroot1988TheJournal.}.

Even in a rare, enigmatic class of objects, P Cygni is unique. It does not exhibit all of the same photometric behavior of other S Doradus variables. It has stayed in its quiescent state, at roughly 18,700 K and 61,000 $L_\sun$, for several hundred years. The nebula surrounding P Cyg is faint and morphologically different from those of most LBVs \citep{Nota1995NebulaePicture}. It has also been studied for longer than any other LBV; an excellent summary of work done in the $20^{th}$ century can be found in \citet{Israelian1999PVariable}. Being such a singular and well studied object, P Cygni is critically important to our understanding of the evolution of the most massive stars in the universe. 

Additionally, since it has been so stable for the past 300 years, ejecta from the outbursts have had the time to move far away from the star, and polarimetry can now open a window into the less dramatic, quiescent mode of mass-loss. There have been extensive studies of the shells of material which were ejected in the 17$^{th}$ century eruptions \citep{Barlow1994TheNebula,Nota1995NebulaePicture,Meaburn1996TheCygni}, first resolved by \citet{Leitherer1987TheImaging.}, but much less is known about the nature of the ambient stellar wind.

In order to study the mass-loss in P Cygni's current state we turn to polarimetry. In the study of massive stars, polarimetry is an invaluable tool. Due to these stars' enormous luminosities and optically thick nebulae, it is difficult to observe the innermost regions of their stellar winds. Polarimetry circumvents this problem, in fact, in certain geometries polarimetric signals can be enhanced in the presence of an optically thick nebula \citep{Wood1996TheEnvelopes,Wood1996TheEnvelopesb}. While P Cygni is in a quiescent (hot) state, we expect electron scattering to be the dominant mechanism causing any net polarization. This scattering is only efficient very near to the star \citep{Nordsieck2001UltravioletCygni,Taylor1991ACygni,Davies2005AsphericityVariables,Davies2006TheWinds} and, as is discussed in the following section, is only sensitive to asymmetries when unresolved. Therefore, polarimetry is a good way to probe asymmetries at the base of P Cygni's stellar wind.

While P Cygni had long been known to be polarized \citep{Coyne1967WavelengthPolarization,Coyne1974WavelengthGrains,Serkowski1975WavelengthExtinction}, the first extensive study of its polarization was conducted by \citet{Hayes1985VariableCygni}. \citet{Taylor1991ACygni} and \citet{Nordsieck2001UltravioletCygni} (hereafter T91 and N01) followed the results of \citet{Hayes1985VariableCygni} using optical spectropolarimetric observations. Spectropolarimetric observations measure the Stokes parameters \textit{q} and \textit{u} as a function of wavelength, just as spectroscopic observations measure flux as a function of wavelength. The following conclusions were made by all three studies \citep{Nordsieck2001UltravioletCygni,Taylor1991ACygni,Hayes1985VariableCygni}:
\begin{itemize}
    \setlength\itemsep{0em}
	\item P Cygni is intrinsically polarized, indicating the presence of asymmetries in the polarizing region.
	\item The polarization of P Cygni does not appear to prefer a position angle. This implies that the asymmetries which produce net polarization are equally likely to emerge in any direction on the plane of the sky.
	\item With previous data sets the polarization of P Cygni did not appear to be periodic, although there was evidence for a 10-15 day characteristic timescale for polarimetric changes. 
\end{itemize}

As presented in the discussion of T91, these results seem to favor a roughly axisymmetric wind with regions of enhanced electron density, referred to as \textit{inhomogeneities} or \textit{clumps}, which are ejected from the star and travel outwards through the polarizing region within its wind. We will use these terms in this work as well.

In this paper, we combine the previously published data from T91 and N01 with unpublished observations from the early 2000s, which were taken with the same instrument as these archival data. In section \ref{sec:obs}, we discuss the details of these observations. We discuss our attempts to remove the polarization due to the interstellar medium in Section \ref{sec:isp}. We analyze the remaining polarization signature, which is intrinsic to the P Cyg system, with novel methods in Section \ref{sec:results}. In Section \ref{sec:dis}, we discuss our results in the larger context of what is already known about the P Cyg system. Finally, we summarize our conclusions in Section \ref{sec:con}.

\section{Observations} \label{sec:obs}
\subsection{HPOL}
Our data are comprised of spectropolarimetric observations taken on 80 nights, spanning 13 years, using the University of Wisconsin's \textit{Half-wave Spectropolarimeter} (HPOL) while it was mounted on the 36'' telescope at Pine Bluff Observatory. In 1995 HPOL switched from using a Reticon dual channel photo-diode array to a $400\times1200$ pixel CCD. As a result our data are split into the 46 observations taken before this switch (hereafter Reticon data) and the 34 taken after (CCD data). The Reticon detector covered a wavelength range between 3200 and 7600 \AA\space at a 15 \AA\space resolution, while the CCD recorded data using blue (3200-6000 \AA) and red (6000-10,500 \AA) gratings in combination to obtain a full observation. The blue grating had a spectral resolution of 7.5 \AA, while the red grating was 10 \AA. More complete reviews of this instrument are in \citet{Nordsieck1996FutureWisconsin}, \citet{Wolff1996ASightlines}, and \citet{Davidson2014TheObservatory}. Basic reduction steps were performed on all of the HPOL observations using the custom built \texttt{Fortran} software package \texttt{REDUCE} \citep{Nook1990TheObservations,Wolff1996ASightlines,Davidson2014TheObservatory}. 

We downloaded these data from the Mikulski Archive for Space Telescopes (MAST) archive\footnote{The website for this archive is https://archive.stsci.edu} and omitted 4 nights of data from our final data set. These four nights of CCD observations only used the red grating (27 September 1997, 27 August 1998, 30 May 2000, and 13 November 2002), which does not cover the full wavelength range we were interested in analyzing. Therefore, we omitted them from our final data set. Table \ref{HPOLObs} lists the civil and modified Julian dates of our remaining HPOL observations (46 Reticon observations and 30 full CCD observations). Of the full 80 nights of data, T91 used 20 of the Reticon observations and N01 used 15 of the CCD observations.

From the normalized Stokes parameters, \textit{q} and \textit{u}, percent polarization, $P_\%$, and position angle, $\Theta$, are defined as
\begin{equation}
    \label{eq:pol}
    P_\% = \sqrt{q^2 + u^2}
\end{equation}

\begin{equation}
    \label{eq:pos}
    \Theta = \frac{1}{2} \arctan{\frac{u}{q}}.
\end{equation}

\subsubsection{Synthetic \textit{V} Band Data} \label{sec:vband}

\begin{figure}[ht!]
    \centering
    \plotone{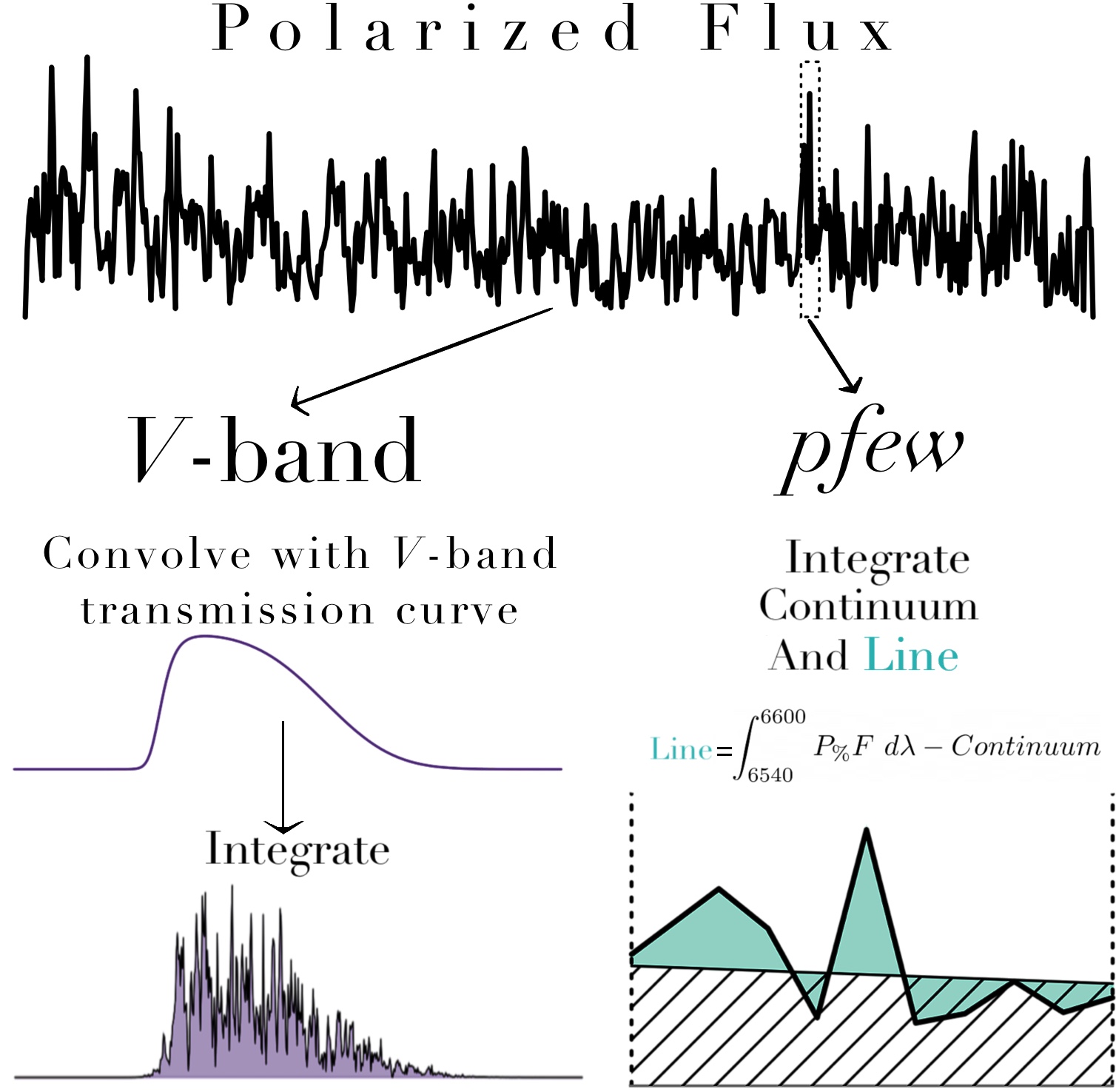}
    \caption{Graphic illustrating how the $V$-band (\S\ref{sec:vband}) and H$\alpha$ $pfew$ (\S\ref{sec:pfew}) values tabulated in Table \ref{HPOLObs} are calculated. These data were taken from a representative observation after ISP correction as described in \S\ref{sec:isp}. For the \textit{pfew} side of the figure (right) the dashed lines represent regions contributing to the continuum polarization and the color represents regions contributing to the line polarization.}
    \label{fig:art}
\end{figure}

To characterize the optical broadband behavior of P Cyg, we convolved each of our HPOL observations with a synthetic Johnson \textit{V}-band filter \citep{Bessell1990UBVRIPassbands}. We report these values in Table \ref{HPOLObs} as our ``observed" V band values. Figure \ref{fig:art} shows a graphical representation of how this convolution process works mathematically.

HPOL's instrumental polarization was monitored on an approximately monthly basis by observing polarized and unpolarized standard stars. \citet{Davidson2014TheObservatory} reported the instrument's systematic uncertainties in the Johnson \textit{UBVRI} bands for the CCD detector, which were calculated using those observations. In order to match HPOL's historically reported values, we calculated one overall root-mean-square systematic uncertainty for each CCD observation in the \textit{V} band using the individual \textit{q} and \textit{u} systematic uncertainties compiled by Davidson et al. (2014). Two observations (10 September 1998 and 16 September 1998) fell outside of all of the date ranges for which \citet{Davidson2014TheObservatory} reported systematic uncertainties. Therefore, we used the values reported for the 14 June 1998-16 August 1998 range because it was closest in time to these observations. Systematic uncertainties for the Reticon detector are less well known, but our previous experience with HPOL data suggests they are less than 0.02\%. In order to characterize the uncertainty in the V band for a given observation, we use the larger of its systematic and internal uncertainties, which are from photon statistics, in our analysis. We report both these uncertainties in Table \ref{HPOLObs}.

\begin{center}
\begin{longrotatetable}
\begin{deluxetable*}{ccc||cc|cc|cc||ccc|ccc|ccc|ccc}
\setlength{\tabcolsep}{2.9pt}
\tablecaption{HPOL Observations \label{HPOLObs}}
\tablehead{
        \colhead{} &\colhead{} &\colhead{} \vline
        &\multicolumn{6}{c}{\textbf{\textit{V} Band}} 
        &\multicolumn{6}{c}{\textbf{H$\alpha$ Continuum}}  
        &\multicolumn{6}{c}{\textbf{H$\alpha$ Line}}  \\
        \colhead{\textbf{Detector}} &\colhead{\textbf{MJD}} &\colhead{\textbf{Date}} \vline
        &\multicolumn{2}{c}{\uline{Observed}} &\multicolumn{2}{c}{\uline{Intrinsic}} &\colhead{} &\colhead{} \vline
        &\multicolumn{3}{c}{\uline{Intrinsic}} &\multicolumn{3}{c}{\uline{Observed}}\vline  &\multicolumn{3}{c}{\uline{Intrinsic}} &\multicolumn{3}{c}{\uline{Observed}} \\
            \colhead{} &\colhead{} &\colhead{} \vline 
            &\colhead{\% q} &\colhead{\% u} \vline &\colhead{\% q} &\colhead{\% u} \vline &\colhead{Int Err} &\colhead{Sys Err} \vline 
            &\colhead{\% q} &\colhead{\% u} &\colhead{Error}\vline 
            &\colhead{\% q} &\colhead{\% u} &\colhead{Error}\vline 
            &\colhead{\% q} &\colhead{\% u} &\colhead{Error}\vline
            &\colhead{\% q} &\colhead{\% u} &\colhead{Error}
}
\startdata
Reticon &  47668 &  1989-05-22 &  0.305 &  0.997 & -0.134 & -0.135 &  0.020 &  0.004 & -0.150 & -0.171 &  0.065 &  0.240 &  0.877 &  0.065 &  0.046 &  0.027 &  0.172 &  0.435 &  1.075 &  0.172 \\
 --- &  47672 &  1989-05-26 &  0.611 &  1.057 &  0.172 & -0.076 &  0.020 &  0.004 &  0.214 & -0.042 &  0.048 &  0.603 &  1.006 &  0.048 & -0.110 &  0.115 &  0.108 &  0.279 &  1.162 &  0.108 \\
 --- &  47697 &  1989-06-20 &  0.289 &  1.128 & -0.150 & -0.005 &  0.020 &  0.004 & -0.058 &  0.042 &  0.053 &  0.332 &  1.090 &  0.053 & -0.227 & -0.143 &  0.101 &  0.163 &  0.905 &  0.101 \\
 --- &  47714 &  1989-07-07 &  0.437 &  0.863 & -0.002 & -0.269 &  0.020 &  0.005 & -0.077 & -0.100 &  0.052 &  0.313 &  0.948 &  0.052 & -0.138 & -0.099 &  0.131 &  0.251 &  0.949 &  0.131 \\
 --- &  47808 &  1989-10-09 &  0.527 &  1.106 &  0.088 & -0.026 &  0.020 &  0.004 &  0.069 & -0.028 &  0.055 &  0.458 &  1.019 &  0.055 & -0.098 & -0.099 &  0.099 &  0.292 &  0.949 &  0.099 \\
 --- &  47822 &  1989-10-23 &  0.174 &  0.826 & -0.265 & -0.306 &  0.020 &  0.003 & -0.193 & -0.269 &  0.037 &  0.197 &  0.779 &  0.037 &  0.028 & -0.168 &  0.064 &  0.417 &  0.879 &  0.064 \\
 --- &  47913 &  1990-01-22 &  0.716 &  1.317 &  0.276 &  0.184 &  0.020 &  0.007 &  0.198 &  0.169 &  0.096 &  0.587 &  1.217 &  0.096 & -0.064 & -0.292 &  0.159 &  0.326 &  0.756 &  0.159 \\
 --- &  47918 &  1990-01-27 &  0.547 &  1.056 &  0.108 & -0.076 &  0.020 &  0.003 &  0.154 & -0.148 &  0.035 &  0.544 &  0.901 &  0.035 &  0.020 & -0.154 &  0.084 &  0.409 &  0.894 &  0.084 \\
 --- &  47919 &  1990-01-28 &  0.675 &  1.111 &  0.236 & -0.021 &  0.020 &  0.003 &  0.163 & -0.051 &  0.040 &  0.553 &  0.997 &  0.040 & -0.027 &  0.060 &  0.079 &  0.362 &  1.108 &  0.079 \\
 --- &  47920 &  1990-01-29 &  0.762 &  1.027 &  0.323 & -0.105 &  0.020 &  0.006 &  0.306 & -0.098 &  0.078 &  0.696 &  0.951 &  0.078 &  0.050 &  0.101 &  0.133 &  0.440 &  1.149 &  0.133 \\
 --- &  47942 &  1990-02-20 &  0.202 &  1.097 & -0.237 & -0.035 &  0.020 &  0.004 & -0.224 & -0.040 &  0.061 &  0.166 &  1.009 &  0.061 & -0.081 & -0.110 &  0.100 &  0.308 &  0.937 &  0.100 \\
 --- &  47950 &  1990-02-28 &  0.353 &  1.387 & -0.086 &  0.255 &  0.020 &  0.003 &  0.009 &  0.275 &  0.044 &  0.399 &  1.323 &  0.044 &  0.010 & -0.055 &  0.091 &  0.399 &  0.992 &  0.091 \\
 --- &  47987 &  1990-04-06 &  0.506 &  1.197 &  0.067 &  0.065 &  0.020 &  0.004 &  0.038 &  0.153 &  0.059 &  0.428 &  1.201 &  0.059 &  0.049 & -0.082 &  0.114 &  0.438 &  0.966 &  0.114 \\
 --- &  47988 &  1990-04-07 &  0.558 &  1.155 &  0.119 &  0.023 &  0.020 &  0.004 &  0.115 &  0.168 &  0.047 &  0.504 &  1.216 &  0.047 & -0.093 & -0.096 &  0.079 &  0.297 &  0.952 &  0.079 \\
 --- &  47994 &  1990-04-13 &  0.355 &  0.955 & -0.084 & -0.178 &  0.020 &  0.003 & -0.086 & -0.119 &  0.031 &  0.303 &  0.929 &  0.031 &  0.007 &  0.028 &  0.064 &  0.397 &  1.075 &  0.064 \\
 --- &  48003 &  1990-04-22 &  0.341 &  1.367 & -0.098 &  0.234 &  0.020 &  0.006 & -0.027 &  0.287 &  0.079 &  0.362 &  1.335 &  0.079 & -0.003 & -0.168 &  0.194 &  0.387 &  0.880 &  0.194 \\
 --- &  48018 &  1990-05-07 &  0.282 &  0.965 & -0.158 & -0.167 &  0.020 &  0.007 & -0.173 & -0.207 &  0.099 &  0.217 &  0.841 &  0.099 &  0.169 & -0.131 &  0.180 &  0.559 &  0.917 &  0.180 \\
 --- &  48047 &  1990-06-05 &  0.556 &  0.977 &  0.116 & -0.155 &  0.020 &  0.015 & -0.028 & -0.278 &  0.248 &  0.362 &  0.770 &  0.248 &  0.250 &  0.168 &  0.446 &  0.640 &  1.215 &  0.446 \\
 --- &  48057 &  1990-06-15 &  0.300 &  1.223 & -0.139 &  0.091 &  0.020 &  0.002 & -0.190 &  0.074 &  0.025 &  0.199 &  1.121 &  0.025 &  0.104 & -0.004 &  0.058 &  0.494 &  1.044 &  0.058 \\
 --- &  48088 &  1990-07-16 &  0.331 &  1.203 & -0.108 &  0.071 &  0.020 &  0.003 & -0.094 &  0.057 &  0.037 &  0.296 &  1.105 &  0.037 & -0.161 & -0.077 &  0.069 &  0.229 &  0.970 &  0.069 \\
 --- &  48133 &  1990-08-30 &  0.759 &  1.077 &  0.320 & -0.056 &  0.020 &  0.003 &  0.193 & -0.060 &  0.039 &  0.583 &  0.988 &  0.039 &  0.201 & -0.121 &  0.067 &  0.590 &  0.926 &  0.067 \\
 --- &  48150 &  1990-09-16 &  0.302 &  1.153 & -0.137 &  0.020 &  0.020 &  0.002 & -0.166 &  0.007 &  0.025 &  0.224 &  1.055 &  0.025 &  0.012 & -0.076 &  0.058 &  0.402 &  0.972 &  0.058 \\
 --- &  48159 &  1990-09-25 &  0.476 &  1.364 &  0.036 &  0.232 &  0.020 &  0.002 & -0.046 &  0.213 &  0.027 &  0.344 &  1.261 &  0.027 &  0.058 & -0.117 &  0.063 &  0.447 &  0.931 &  0.063 \\
 --- &  48224 &  1990-11-29 &  0.472 &  0.962 &  0.033 & -0.170 &  0.020 &  0.002 &  0.068 & -0.123 &  0.023 &  0.458 &  0.925 &  0.023 & -0.081 & -0.114 &  0.063 &  0.309 &  0.935 &  0.063 \\
 --- &  48231 &  1990-12-06 &  0.275 &  0.495 & -0.164 & -0.637 &  0.020 &  0.002 & -0.118 & -0.636 &  0.023 &  0.271 &  0.412 &  0.023 & -0.066 &  0.089 &  0.047 &  0.324 &  1.137 &  0.047 \\
 --- &  48235 &  1990-12-10 &  0.099 &  1.225 & -0.340 &  0.092 &  0.020 &  0.002 & -0.329 &  0.052 &  0.020 &  0.061 &  1.100 &  0.020 &  0.085 & -0.225 &  0.046 &  0.474 &  0.822 &  0.046 \\
 Reticon &  48320 &  1991-03-05 &  0.279 &  1.290 & -0.161 &  0.158 &  0.020 &  0.003 & -0.171 &  0.117 &  0.046 &  0.219 &  1.165 &  0.046 &  0.002 & -0.090 &  0.088 &  0.392 &  0.958 &  0.088 \\
 --- &  48444 &  1991-07-07 &  0.523 &  1.286 &  0.084 &  0.154 &  0.020 &  0.005 &  0.133 &  0.046 &  0.068 &  0.523 &  1.094 &  0.068 & -0.086 &  0.007 &  0.128 &  0.303 &  1.055 &  0.128 \\
 --- &  48473 &  1991-08-05 &  0.537 &  0.973 &  0.098 & -0.159 &  0.020 &  0.014 &  0.348 & -0.028 &  0.264 &  0.738 &  1.020 &  0.264 &  0.036 &  0.094 &  0.360 &  0.426 &  1.141 &  0.360 \\
 --- &  48516 &  1991-09-17 &  0.596 &  1.150 &  0.156 &  0.018 &  0.020 &  0.002 &  0.128 &  0.005 &  0.028 &  0.518 &  1.053 &  0.028 & -0.034 &  0.032 &  0.045 &  0.355 &  1.080 &  0.045 \\
 --- &  48520 &  1991-09-21 &  0.135 &  0.817 & -0.305 & -0.315 &  0.020 &  0.002 & -0.275 & -0.369 &  0.027 &  0.115 &  0.679 &  0.027 &  0.022 & -0.032 &  0.046 &  0.411 &  1.016 &  0.046 \\
 --- &  48522 &  1991-09-23 &  0.024 &  1.021 & -0.415 & -0.111 &  0.020 &  0.002 & -0.389 & -0.187 &  0.025 &  0.001 &  0.861 &  0.025 &  0.016 & -0.142 &  0.040 &  0.406 &  0.906 &  0.040 \\
 --- &  48525 &  1991-09-26 &  0.799 &  1.336 &  0.360 &  0.203 &  0.020 &  0.004 &  0.277 &  0.197 &  0.060 &  0.666 &  1.245 &  0.060 & -0.122 & -0.036 &  0.098 &  0.268 &  1.012 &  0.098 \\
 --- &  48526 &  1991-09-27 &  0.843 &  1.348 &  0.404 &  0.216 &  0.020 &  0.002 &  0.429 &  0.112 &  0.036 &  0.819 &  1.160 &  0.036 & -0.137 & -0.010 &  0.053 &  0.253 &  1.038 &  0.053 \\
 --- &  48527 &  1991-09-28 &  0.796 &  1.226 &  0.357 &  0.094 &  0.020 &  0.002 &  0.367 &  0.062 &  0.025 &  0.757 &  1.110 &  0.025 & -0.111 & -0.085 &  0.040 &  0.278 &  0.963 &  0.040 \\
 --- &  48528 &  1991-09-29 &  0.740 &  1.089 &  0.301 & -0.044 &  0.020 &  0.002 &  0.295 & -0.046 &  0.027 &  0.685 &  1.002 &  0.027 & -0.039 &  0.058 &  0.041 &  0.351 &  1.105 &  0.041 \\
 --- &  48529 &  1991-09-30 &  0.659 &  1.038 &  0.220 & -0.094 &  0.020 &  0.003 &  0.171 & -0.148 &  0.030 &  0.560 &  0.900 &  0.030 & -0.043 &  0.006 &  0.055 &  0.346 &  1.053 &  0.055 \\
 --- &  48530 &  1991-10-01 &  0.669 &  1.047 &  0.230 & -0.086 &  0.020 &  0.002 &  0.195 & -0.108 &  0.030 &  0.585 &  0.940 &  0.030 & -0.017 & -0.109 &  0.051 &  0.372 &  0.938 &  0.051 \\
 --- &  49521 &  1994-06-18 &  0.432 &  0.732 & -0.007 & -0.400 &  0.020 &  0.008 & -0.093 & -0.368 &  0.103 &  0.296 &  0.680 &  0.103 &  0.092 & -0.205 &  0.201 &  0.481 &  0.843 &  0.201 \\
 --- &  49530 &  1994-06-27 &  0.551 &  0.790 &  0.112 & -0.342 &  0.020 &  0.005 &  0.077 & -0.236 &  0.064 &  0.467 &  0.812 &  0.064 &  0.169 & -0.273 &  0.299 &  0.558 &  0.775 &  0.299 \\
 --- &  49539 &  1994-07-06 &  0.770 &  0.694 &  0.331 & -0.438 &  0.020 &  0.011 &  0.232 & -0.491 &  0.127 &  0.621 &  0.557 &  0.127 &  0.191 &  0.325 &  0.684 &  0.580 &  1.373 &  0.684 \\
 --- &  49573 &  1994-08-09 &  0.222 &  0.974 & -0.217 & -0.158 &  0.020 &  0.006 & -0.134 & -0.082 &  0.088 &  0.256 &  0.965 &  0.088 &  0.019 & -0.248 &  0.154 &  0.409 &  0.799 &  0.154 \\
 --- &  49606 &  1994-09-11 &  0.156 &  1.221 & -0.283 &  0.089 &  0.020 &  0.004 & -0.281 &  0.106 &  0.067 &  0.109 &  1.154 &  0.067 &  0.137 & -0.188 &  0.156 &  0.527 &  0.860 &  0.156 \\
 --- &  49636 &  1994-10-11 &  0.208 &  1.059 & -0.232 & -0.074 &  0.020 &  0.003 & -0.307 & -0.056 &  0.041 &  0.083 &  0.992 &  0.041 & -0.077 & -0.158 &  0.101 &  0.312 &  0.890 &  0.101 \\
 --- &  49666 &  1994-11-10 &  0.159 &  1.047 & -0.280 & -0.085 &  0.020 &  0.003 & -0.282 & -0.066 &  0.041 &  0.108 &  0.982 &  0.041 & -0.003 & -0.263 &  0.133 &  0.387 &  0.784 &  0.133 \\
 --- &  49690 &  1994-12-04 &  0.142 &  0.901 & -0.297 & -0.231 &  0.020 &  0.004 & -0.352 & -0.195 &  0.054 &  0.038 &  0.853 &  0.054 &  0.068 & -0.227 &  0.134 &  0.457 &  0.820 &  0.134 \\
 CCD &  49777 &  1995-03-01 &  0.291 &  1.241 & -0.125 &  0.121 &  0.006 &  0.005 & -0.084 &  0.170 &  0.032 &  0.304 &  1.213 &  0.032 &  0.037 &  0.002 &  0.058 &  0.427 &  1.053 &  0.058 \\
 --- &  49779 &  1995-03-03 &  0.308 &  1.140 & -0.109 &  0.018 &  0.006 &  0.003 & -0.086 &  0.085 &  0.018 &  0.301 &  1.127 &  0.018 & -0.044 & -0.035 &  0.028 &  0.346 &  1.015 &  0.028 \\
 --- &  49785 &  1995-03-09 &  0.051 &  1.169 & -0.366 &  0.048 &  0.006 &  0.002 & -0.358 &  0.041 &  0.019 &  0.029 &  1.084 &  0.019 & -0.093 &  0.051 &  0.041 &  0.298 &  1.102 &  0.041 \\
 --- &  49788 &  1995-03-12 &  0.285 &  1.577 & -0.132 &  0.455 &  0.006 &  0.003 & -0.114 &  0.363 &  0.035 &  0.273 &  1.406 &  0.035 & -0.085 & -0.041 &  0.053 &  0.305 &  1.010 &  0.053 \\
 --- &  49891 &  1995-06-23 &  0.600 &  1.110 &  0.184 & -0.010 &  0.006 &  0.003 &  0.226 &  0.049 &  0.021 &  0.614 &  1.092 &  0.021 & -0.026 & -0.025 &  0.067 &  0.365 &  1.027 &  0.067 \\
 --- &  49913 &  1995-07-15 &  0.339 &  1.247 & -0.077 &  0.127 &  0.010 &  0.003 & -0.004 &  0.145 &  0.017 &  0.384 &  1.188 &  0.017 &  0.012 & -0.059 &  0.042 &  0.403 &  0.993 &  0.042 \\
 --- &  49940 &  1995-08-11 &  0.283 &  1.055 & -0.133 & -0.066 &  0.010 &  0.003 & -0.072 &  0.023 &  0.017 &  0.316 &  1.066 &  0.017 & -0.017 &  0.021 &  0.050 &  0.375 &  1.074 &  0.050 \\
 --- &  49965 &  1995-09-05 &  0.482 &  1.381 &  0.066 &  0.261 &  0.010 &  0.003 &  0.030 &  0.249 &  0.019 &  0.418 &  1.293 &  0.019 & -0.063 & -0.032 &  0.076 &  0.328 &  1.019 &  0.076 \\
 --- &  49977 &  1995-09-17 &  0.666 &  0.961 &  0.249 & -0.160 &  0.010 &  0.002 &  0.262 & -0.226 &  0.017 &  0.650 &  0.817 &  0.017 & -0.138 & -0.007 &  0.050 &  0.254 &  1.047 &  0.050 \\
 CCD &  49992 &  1995-10-02 &  0.210 &  1.049 & -0.207 & -0.071 &  0.010 &  0.002 & -0.192 & -0.098 &  0.013 &  0.196 &  0.944 &  0.013 & -0.024 &  0.053 &  0.041 &  0.368 &  1.108 &  0.041 \\
 --- &  50009 &  1995-10-19 &  0.242 &  1.008 & -0.174 & -0.112 &  0.010 &  0.004 & -0.158 & -0.052 &  0.018 &  0.230 &  0.991 &  0.018 &  0.010 &  0.014 &  0.052 &  0.402 &  1.068 &  0.052 \\
 --- &  50036 &  1995-11-15 &  0.382 &  1.181 & -0.034 &  0.060 &  0.010 &  0.002 & -0.018 &  0.051 &  0.014 &  0.370 &  1.094 &  0.014 & -0.083 & -0.003 &  0.051 &  0.309 &  1.051 &  0.051 \\
 --- &  50055 &  1995-12-04 &  0.289 &  1.222 & -0.127 &  0.101 &  0.010 &  0.003 & -0.132 &  0.148 &  0.016 &  0.256 &  1.190 &  0.016 & -0.057 & -0.089 &  0.061 &  0.334 &  0.964 &  0.061 \\
 --- &  50733 &  1997-10-12 &  0.312 &  1.077 & -0.105 & -0.043 &  0.009 &  0.003 & -0.094 & -0.098 &  0.014 &  0.293 &  0.944 &  0.014 &  0.165 &  0.193 &  0.084 &  0.562 &  1.259 &  0.084 \\
 --- &  50803 &  1997-12-21 &  0.476 &  1.173 &  0.059 &  0.052 &  0.009 &  0.002 &  0.074 &  0.070 &  0.014 &  0.461 &  1.111 &  0.014 & -0.037 & -0.304 &  0.073 &  0.359 &  0.760 &  0.073 \\
 --- &  51010 &  1998-07-16 &  0.411 &  1.045 & -0.006 & -0.075 &  0.010 &  0.002 & -0.015 & -0.106 &  0.013 &  0.372 &  0.936 &  0.013 & -0.093 & -0.022 &  0.068 &  0.302 &  1.040 &  0.068 \\
 --- &  51034 &  1998-08-09 &  0.711 &  0.966 &  0.294 & -0.155 &  0.010 &  0.003 &  0.332 & -0.146 &  0.015 &  0.719 &  0.896 &  0.015 & -0.323 &  0.256 &  0.075 &  0.072 &  1.319 &  0.075 \\
 --- &  51066 &  1998-09-10 &  0.470 &  1.170 &  0.053 &  0.049 &  0.010 &  0.002 &  0.072 &  0.018 &  0.013 &  0.460 &  1.060 &  0.013 &  0.043 &  0.151 &  0.067 &  0.438 &  1.211 &  0.067 \\
 --- &  51072 &  1998-09-16 &  0.390 &  1.203 & -0.027 &  0.082 &  0.010 &  0.002 &  0.018 &  0.073 &  0.015 &  0.405 &  1.115 &  0.015 &  0.023 & -0.127 &  0.067 &  0.418 &  0.934 &  0.067 \\
 --- &  51160 &  1998-12-13 &  0.518 &  1.277 &  0.101 &  0.156 &  0.008 &  0.003 &  0.079 &  0.150 &  0.014 &  0.467 &  1.193 &  0.014 & -0.024 & -0.052 &  0.048 &  0.369 &  1.002 &  0.048 \\
 --- &  51177 &  1998-12-30 &  0.265 &  1.069 & -0.152 & -0.052 &  0.008 &  0.004 & -0.171 & -0.034 &  0.028 &  0.217 &  1.009 &  0.028 & -0.012 &  0.085 &  0.059 &  0.378 &  1.136 &  0.059 \\
 --- &  52533 &  2002-09-16 &  0.210 &  0.970 & -0.206 & -0.151 &  0.012 &  0.004 & -0.219 & -0.067 &  0.024 &  0.169 &  0.976 &  0.024 & -0.151 & -0.012 &  0.050 &  0.239 &  1.039 &  0.050 \\
 --- &  52534 &  2002-09-17 &  0.089 &  0.949 & -0.328 & -0.171 &  0.012 &  0.004 & -0.367 & -0.111 &  0.030 &  0.021 &  0.932 &  0.030 &  0.126 &  0.056 &  0.056 &  0.517 &  1.107 &  0.056 \\
 --- &  52541 &  2002-09-24 &  0.236 &  1.029 & -0.181 & -0.091 &  0.012 &  0.004 & -0.219 & -0.047 &  0.027 &  0.169 &  0.996 &  0.027 & -0.153 & -0.095 &  0.058 &  0.238 &  0.956 &  0.058 \\
 --- &  52542 &  2002-09-25 &  0.237 &  1.031 & -0.179 & -0.090 &  0.012 &  0.006 & -0.248 & -0.033 &  0.030 &  0.140 &  1.010 &  0.030 & -0.053 & -0.018 &  0.075 &  0.338 &  1.034 &  0.075 \\
 --- &  52561 &  2002-10-14 &  0.589 &  1.035 &  0.172 & -0.085 &  0.012 &  0.004 &  0.150 & -0.078 &  0.022 &  0.538 &  0.965 &  0.022 & -0.181 & -0.005 &  0.056 &  0.210 &  1.047 &  0.056 \\
 --- &  52562 &  2002-10-15 &  0.487 &  1.117 &  0.070 & -0.003 &  0.012 &  0.004 &  0.096 &  0.014 &  0.022 &  0.483 &  1.056 &  0.022 & -0.078 & -0.072 &  0.048 &  0.312 &  0.979 &  0.048 \\
 --- &  52579 &  2002-11-01 &  0.293 &  0.919 & -0.123 & -0.201 &  0.012 &  0.004 & -0.121 & -0.137 &  0.025 &  0.267 &  0.906 &  0.025 & -0.070 & -0.112 &  0.055 &  0.321 &  0.939 &  0.055 \\
 --- &  52583 &  2002-11-05 &  0.176 &  0.953 & -0.241 & -0.168 &  0.012 &  0.005 & -0.281 & -0.170 &  0.079 &  0.107 &  0.873 &  0.079 & -0.082 &  0.031 &  0.147 &  0.309 &  1.083 &  0.147 \\
 --- &  52586 &  2002-11-08 &  0.449 &  0.839 &  0.032 & -0.282 &  0.012 &  0.004 & -0.081 & -0.204 &  0.025 &  0.307 &  0.839 &  0.025 &  0.023 & -0.134 &  0.043 &  0.414 &  0.916 &  0.043 
\enddata
\end{deluxetable*}
\end{longrotatetable}
\end{center}

\subsubsection{H$\alpha$: the \textit{pfew} Method} \label{sec:pfew}
In an attempt to quantify the overall behavior of the H$\alpha$ emission line, we calculated its polarized flux equivalent width (\textit{pfew}) \citep{Lomax2014,Hoffman1998}, which integrates the Stokes $q$ and $u$ values in the emission line and removes underlying continuum polarization (see Figure \ref{fig:art} for a graphical representation of the mathematics used the \textit{pfew} method). These values are reported in Table \ref{HPOLObs}. In order to do this, we identified two continuum regions on either side of the line that the \textit{pfew} method uses to determine the line's underlying continuum: 6400 to 6475 \AA\space on the blue side and 6750 to 6800 \AA\space on the red side. Additionally, we integrated over the line between 6540 and 6600 \AA. 

In the remainder of this work we use the term \textit{pfew} to refer to both this method and the values it produced---listed under H$\alpha$ Continuum and H$\alpha$ Line columns in Table \ref{HPOLObs}.

\subsection{WUPPE}
We also used three ultraviolet (UV) spectropolarimetric observations of P Cyg taken with the Wisconsin UV Photo-Polarimeter Experiment (WUPPE). WUPPE had a resolution of approximately 12 \AA\space and recorded spectropolarimetric data between 1400 and 3200 \AA. P Cyg was observed on four separate occasions with WUPPE, which flew aboard both the Space Shuttle Columbia as part of the STS-35 ASTRO-1 mission and the Space Shuttle Endeavour as part of the STS-67 ASTRO-2 mission.  More information about this instrument can be found in \citet{Bjorkman1993UltravioletExperiment} and \citet{Nordsieck1994titleExploringWUPPE/title}.

The ASTRO-1 data were taken on 5 December 1990 and originally published in \citet{Taylor1991FirstSupergiants}. The three ASTRO-2 observations, taken on 3, 8, and 12 March 1995, show little to no variability suggesting P Cyg has a constant UV polarimetric behavior. Therefore, we did not include the ASTRO-1 observation in our analysis.

\section{Interstellar Polarization} \label{sec:isp}
\begin{figure}
    \centering
    \plotone{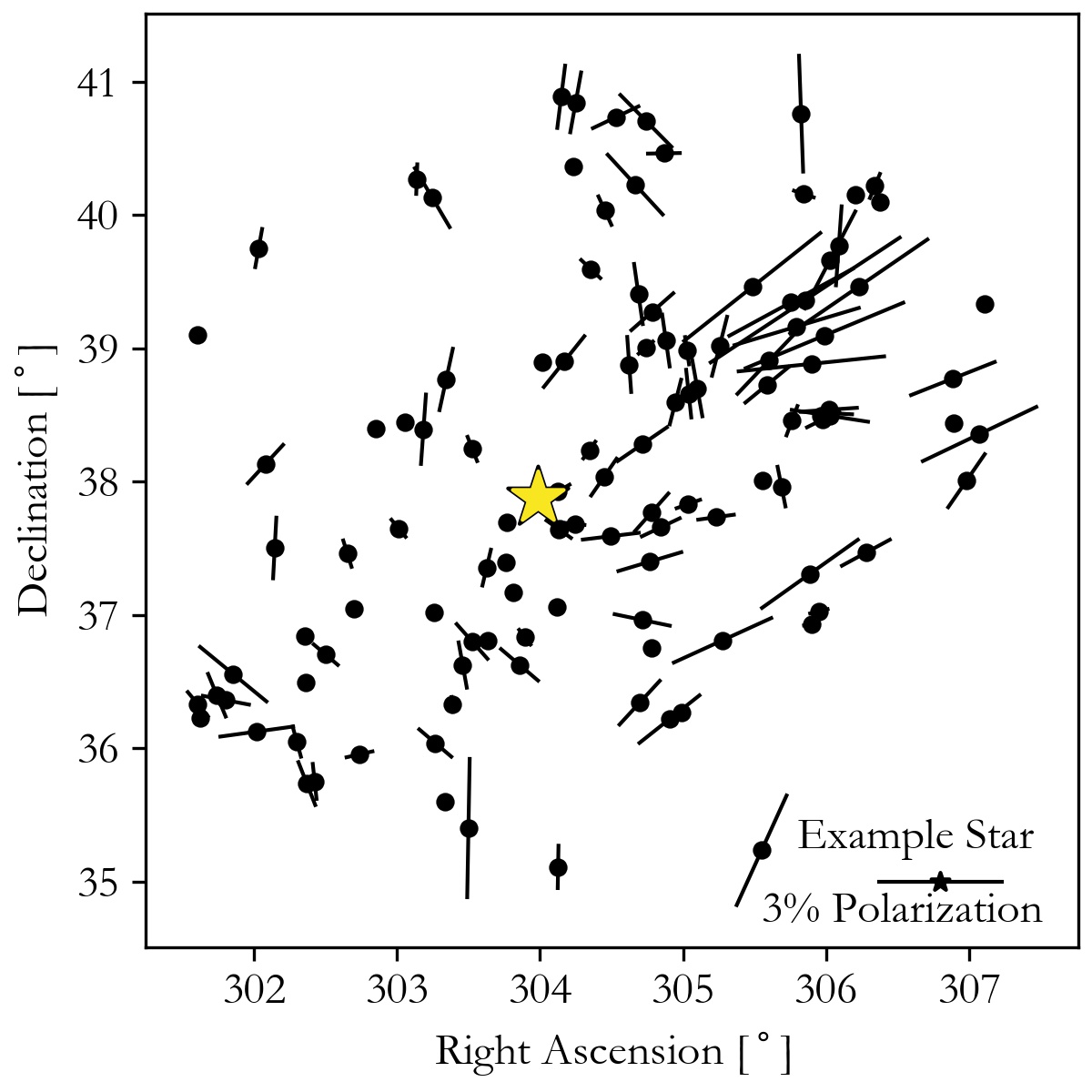}
   \caption{The stars within $\sim3^\circ$ of P Cygni found in \citet{Heiles19999286Catalogs}. The length of the line going through the position of each star represents the relative strength of the star’s polarization and the orientation of the line represents position angle. The yellow star shows the position of P Cygni.}
    \label{fig:map}
\end{figure}
In general, we expect the light coming from spatially unresolved stars to be unpolarized. It only becomes polarized when those photons scatter off of something, which is typically either circumstellar dust around a star, free electrons in the circumstellar medium (CSM) around a star, or dust in the interstellar medium. Because P Cyg is a hot star, its circumstellar material is highly ionized. Therefore, the polarization we observe is from a combination of electron scattering within P Cyg's CSM and scattering with dust in the interstellar medium between P Cyg and Earth. Before analyzing our spectropolarimetric observations of P Cyg, we separate polarization which is intrinsic to the P Cyg system due to electron scattering in the CSM (hereafter called intrinsic polarization) from that which is produced by the interstellar medium (hereafter called interstellar polarization or ISP).

\begin{figure}
    \plotone{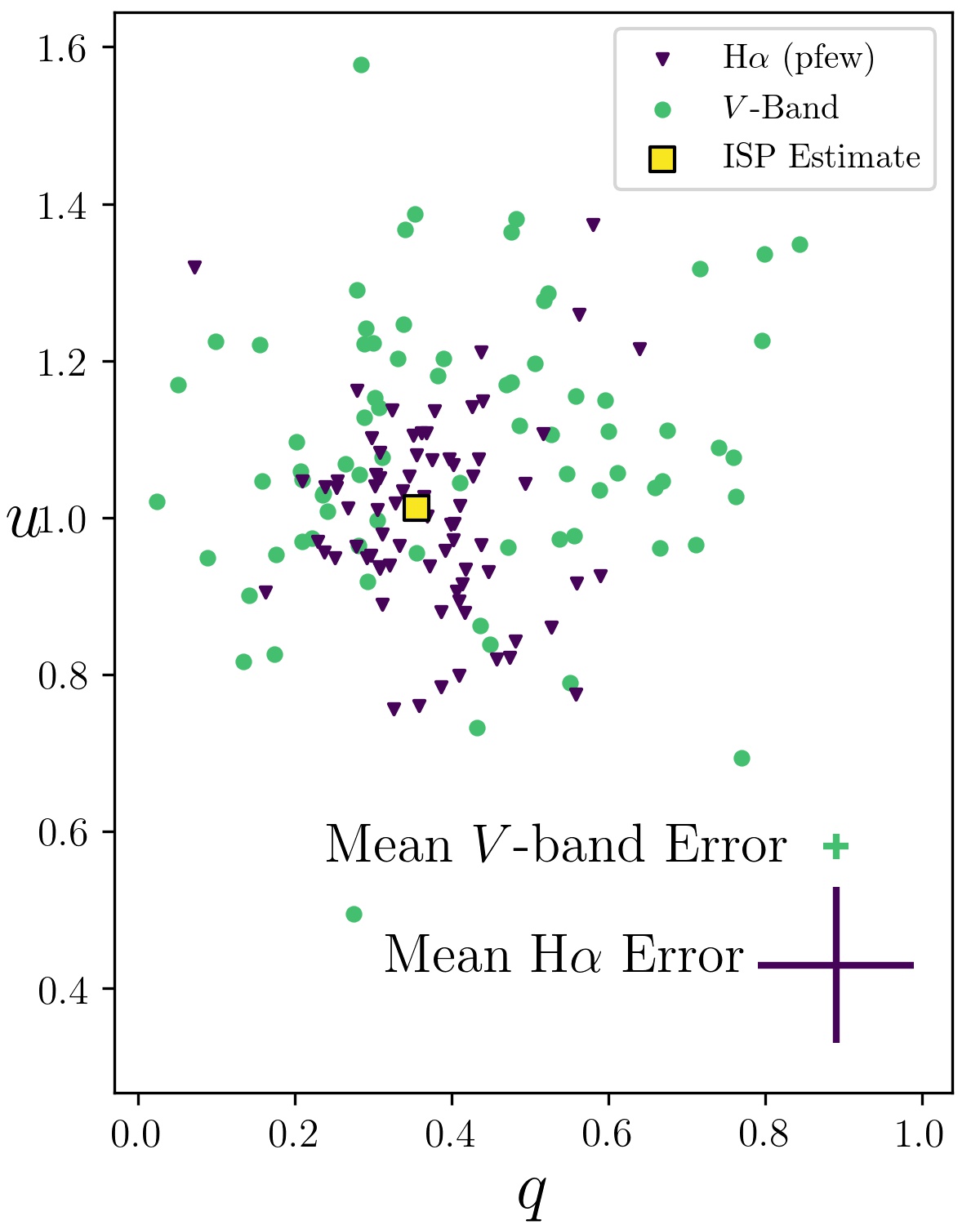}
    \caption{76 synthetic $V$-band \textit{HPOL} observations (see \S\ref{sec:vband}), observed H$\alpha$ line polarization (see \S\ref{sec:pfew}), and our estimate of the ISP at H$\alpha$ (see \S\ref{sec:isp}; Table \ref{HPOLObs}).}
    \label{fig:comp}
\end{figure}

Electron scattering is a gray process that produces a constant polarization signal in both \textit{q} and \textit{u} (or \textit{$P_\%$} and $\Theta$) with wavelength. However, a polarization signal from electron scattering can change with time if the geometry of the scattering region (i.e. region where the free electrons are located) is changing or the number of free electrons for photons to scatter off of changes. Therefore, the polarization signal due to electron scattering from P Cyg may change location in the $q$ vs. $u$ plane (i.e. Figure \ref{fig:comp}) with time, but the polarization signal at different wavelengths will lie on the same point if they were taken simultaneously.

Conversely, the ISP has a wavelength dependence, but is generally thought to be constant over long periods of time. In the $q$ vs. $u$ plane (i.e. Figure \ref{fig:comp}) the interstellar polarization signal manifests itself as a vector that changes with wavelength which adds to the constant electron scattering signal from P Cyg. In particular, it is the $P_\%$ of the ISP that changes with wavelength, while the $\Theta$ of the ISP is constant with wavelength \citep{Serkowski1975WavelengthExtinction,Wilking1982TheLambda/max/}. Thus one must construct a model of the wavelength dependence of the $P_\%$ of the ISP along the line of sight to P Cygni---which is caused by light scattering off of dust grains in the interstellar medium---to subtract from the observed polarization in order to determine P Cygni's intrinsic polarization. The wavelength dependence of the ISP can be modeled with the Serkowski law \citep{Serkowski1975WavelengthExtinction}, as modified by \citet{Wilking1982TheLambda/max/}:
\begin{equation}
    P_\lambda = P_{max} e^{-K \ln^2{\lambda_{max}/\lambda}}
    \label{eq:wrl}
\end{equation}
where $P_{max}$ is the largest value of $P_\%$ due to the ISP, $\lambda_{max}$ is the wavelength at which that peak occurs, and,
$$K = (0.01\pm0.05) + (1.66\pm0.09)\lambda_{max},$$ where $\lambda_{max}$ is measured in microns \citep{Whittet1992SystematicPolarization}.

Therefore, to determine the Serkowski law in the direction of P Cygni, we must find:
\begin{enumerate}
    \setlength\itemsep{0em}
    \item at least one wavelength at which we know the magnitude of the ISP---This allows us to solve for $P_{max}$ in equation \ref{eq:wrl}.
    \item the wavelength of peak polarization, $\lambda_{max}$
\end{enumerate}
A summary of techniques and challenges relating to determining the ISP can be found in \citet{Wisniewski2010Disk-lossData}.

Traditionally, the most common method to find a known ISP value is to observe stars near the science target in the sky and find the average direction and strength of their polarization and take this to be the value of the ISP. Unfortunately, the Cygnus region is notoriously complex. This is illustrated in Figure \ref{fig:map}, which shows the relative strength and orientation of linear polarization for the stars in the \citet{Heiles19999286Catalogs} polarimetric catalog within $\sim3^\circ$ of P Cygni. Taking only the subset of stars at a similar distance as P Cygni does not help to establish a pattern. It is worth noting that the \citet{Heiles19999286Catalogs} catalogue does not provide the wavelength at which polarization was measured. However, since we do not expect a significant wavelength dependence in the observed position angle of the ISP, it is generally still valuable to compare nearby stars in this manner. In the case of P Cygni, the large variations in the position angles of its surrounding field stars (Figure \ref{fig:map}) make this method of determining the ISP less than ideal.

\begin{figure*}[ht!]
    \centering
    \plotone{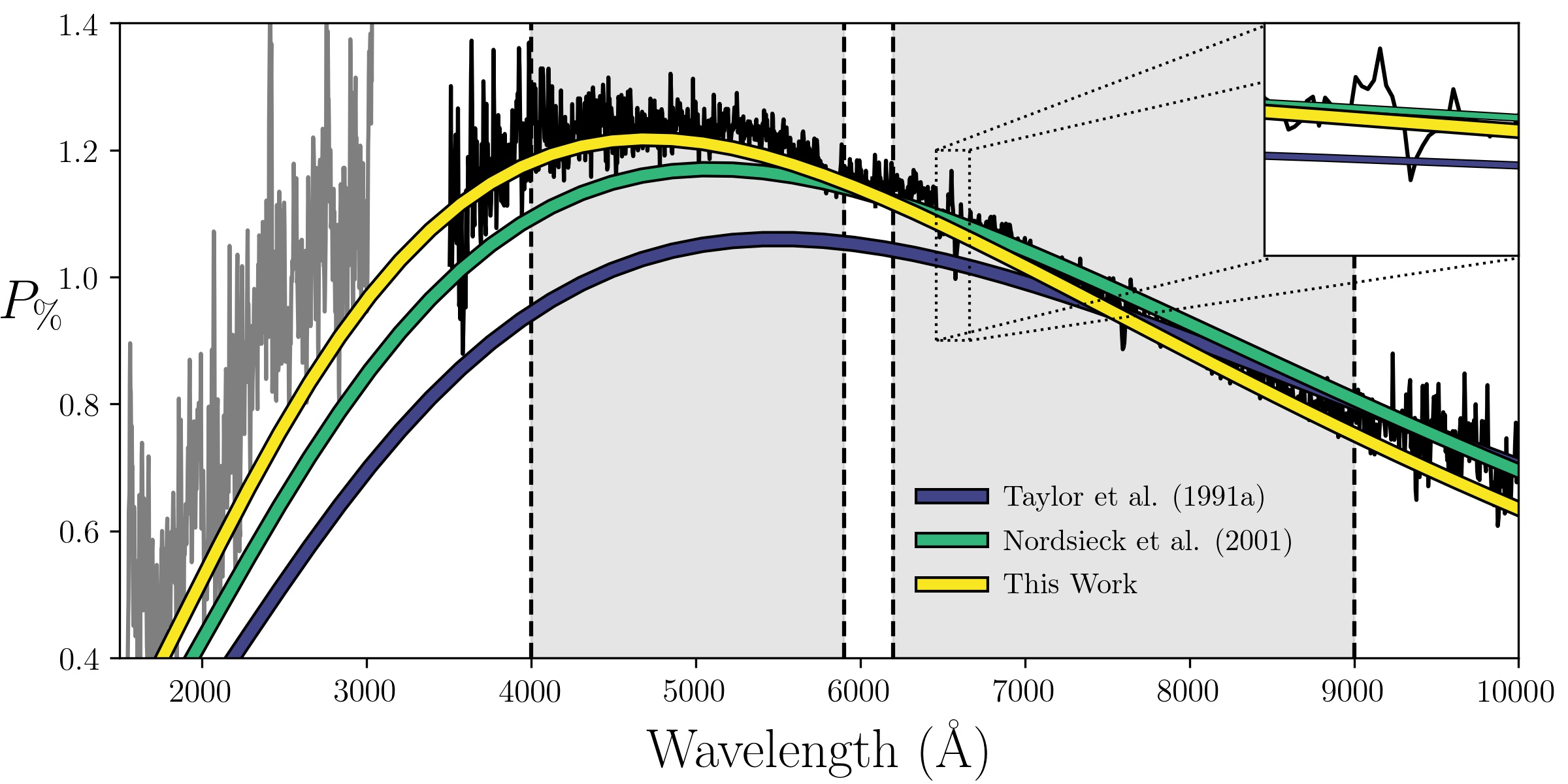}
    \caption{Comparison of ISP estimates from \citet{Nordsieck2001UltravioletCygni}, \citet{Taylor1991ACygni}, and this paper. These models are plotted over the mean HPOL CCD data in black (see Section 3), and the mean of 3 UV observations from the WUPPE polarimeter in grey. The zoom-in in the upper-righthand corner shows the region around H$\alpha$ in greater detail.}
    \label{fig:isp}
\end{figure*}

Alternatively, the \textit{line center method} makes the assumption that strong emission lines, such as H$\alpha$, are intrinsically unpolarized. In P Cygni, these emission lines are formed between 3.2 and 9.3 $R_\star$ \citep{Avcioglu1984StructureCygni}, beyond the region where free-electron scattering is efficient; i.e. the H$\alpha$ line forms farther away from the star than the region where electron scattering occurs \citep{Nordsieck2001UltravioletCygni,Taylor1991ACygni,Davies2005AsphericityVariables,Davies2006TheWinds}. Therefore, the observed polarization of these lines are likely a direct measurement of the ISP, which we do not expect to vary on the timescales over which our data were taken. We find evidence for this in Figure \ref{fig:comp} which shows the distribution of the 76 synthetic $V$-band observations and measurements of the polarization in the H$\alpha$ emission line (\S\ref{sec:pfew}) in $q$-$u$ space. It can be seen that H$\alpha$ varies less than the $V$-band observations ($\sigma_V \simeq 0.2$ while $\sigma_H \simeq 0.1$), suggesting it is less affected by a time-variable electron scattering component (i.e. polarization intrinsic to P Cyg). This makes H$\alpha$ a better estimator of the ISP, which likely lies near the center of both distributions.

Both T91 and N01 presented models for the ISP, constructed using different methods and data. To estimate $\lambda_{max}$, T91 made use of the relationship between the ratio of total-to-selective extinction, $R_V$, and $\lambda_{max}$ in \citet{Serkowski1975WavelengthExtinction}:
\begin{equation}
    R_V = (5.6 \pm 0.3)\lambda_{max}
    \label{eq:rv}
\end{equation}
and assumed the galactic average, $R_v = 3.1$ \citep{Cardelli1989TheExtinction}, to find a peak polarization wavelength of 5500 \AA{}. They then employed the \textit{line center method} and assumed that H$\alpha$ is intrinsically unpolarized to estimate $P_{max}$ (1.06\%) using their spectropolarimetric observations of P Cygni. With both $P_{max}$ and $\lambda_{max}$ they were able to construct a model of the ISP in the direction of P Cygni.

N01 followed a method to similar to T91's to estimate the ISP. They also assumed the H$\alpha$ was intrinsically unpolarized and estimated the ISP at that wavelength. They used UV data from the Wisconsin Ultraviolet Photo-Polarimeter Experiment (WUPPE) and assumed that the region between 1700-1900 \AA{} is also unpolarized to get an estimate of the ISP at those wavelengths. With the combination of those two data points (the H$\alpha$ and UV ISP) they were able to use the Serkowski law to back out both a $P_{max}$ ($1.17 \pm 0.03\% $) and a $\lambda_{max}$ (5100 \AA{}).

When we subtracted the ISP estimates from T91 and N01 from our HPOL data, we found that the resulting polarization was strongly wavelength dependent for most observations. Because P Cygni is a hot star, we expect that its intrinsic polarization is due to electron scattering, which is a `gray' process that should have no wavelength dependence. Therefore, we determined our own ISP estimate to try to resolve this issue. 

Using the \textit{line center method} (despite the problems discussed in \S\ref{sec:dis}) we initially performed a similar analysis, using $pfew$ measurements of H$\alpha$, eq. \ref{eq:rv} and $R_v = 3.1$. However, we found that the peak wavelength which best describes our overall dataset is bluer than both T91's and N01's previously used $\lambda_{max}$ estimates (5500\AA{} and 5100\AA{}). However, both T91 and N01 also noted that they observed a persistent decrease of P Cygni's intrinsic polarization into the infrared whose strength varies, suggesting there might be a high amount of free-free absorptive opacity due to clumps at the base of the system's wind. 

To more rigorously determine $\lambda_{max}$ instead of assuming that $R_V = 3.1$, we first found the mean $P_\%$ curve of all our CCD observations. We calculated this by finding the error-weighted mean Stokes \textit{q} and \textit{u} parameters at every wavelength for which we have CCD data and then combined those mean Stokes parameters into $P_\%$ (Figure \ref{fig:isp}). We then performed a least squares fit, using equation \ref{eq:wrl} and \texttt{SciPy}'s \textit{curve}\_\textit{fit} function \citep{Jones2001SciPy:Python}, to two sections of our mean $P_\%$ curve (between 4000 and 5900 \AA{} and between 6200 and 9000 \AA{}) simultaneously, which were chosen to avoid including data that fell near the edges of the CCD detector that typically have higher uncertainties. As part of this process, we allowed $P_{max}$ and $\lambda_{max}$ in equation \ref{eq:wrl} to vary and found $\lambda_{max} = 4712 \pm 51$ \AA.

We then estimated the $P_{\%}$ at H$\alpha$ because that should only be due to the ISP. We used the \textit{pfew} H$\alpha$ line values tabulated above, found the mean H$\alpha$ Stokes \textit{q} and \textit{u} value weighted inversely by variance, and converted that into a mean $P_{\%}$ at H$\alpha$. We used this value, equation \ref{eq:wrl}, and our previously determined $\lambda_{max}$ (4712\AA) to find $P_{\%,ISP} = 1.06\%$. Because the ISP is not expected to have a wavelength dependence in its position angle, we found the average position angle of all of our observations and assumed that was representative of the ISP ($\Theta_{ISP}=34.4^\circ \pm 0.7^\circ$).

We then use equation \ref{eq:rv} to solve for $R_V$ in the direction of P Cygni. Given our value of $\lambda_{max}$ and equation \ref{eq:rv}, we find that $R_V = 2.638 \pm 0.028$. This lower, when compared to T91, value of $R_V$---corresponding to a smaller size of dust grains in the interstellar medium---is in excellent agreement with the value of $R_V = 2.73 \pm 0.23$ determined by \citet{Turner2001CharacteristicsMembership}. This result can be explained by the dust being processed by ionizing radiation from P Cygni and other massive stars in the OB association \citep{Draine2003}, and suggests we have a robust ISP estimate to subtract from our observations. Our resulting ISP fit is plotted in yellow in Figure \ref{fig:isp}. We also over plot the T91 and N01 ISP estimates for comparison.

To be clear, our ISP model itself is \textit{not} a fit to the data. The fitting done above is only for determining $\lambda_{max}$. Using $\lambda_{max}$ and the polarization at H$\alpha$ (which we assume to be the same as the value of the ISP at H$\alpha$) we \textit{solve} for $P_{max}$. Plugging $\lambda_{max}$ and $P_{max}$ into equation \ref{eq:wrl} yeilds the wavelength dependence of the ISP signal.

It is worth noting that even though we do not use the WUPPE data, our ISP estimate agrees well with N01's assumption that the 1700-1900 \AA{} region of P Cygni is intrinsically unpolarized (i.e. our estimate overlaps with the WUPPE data in this wavelength regime). The small vertical offset from our ISP fit can be explained by P Cygni being a ``super-Serkowski'' object as discussed in \citet{Anderson1996UltravioletExperiment}.

\begin{figure}[!ht]
    \centering
    \plotone{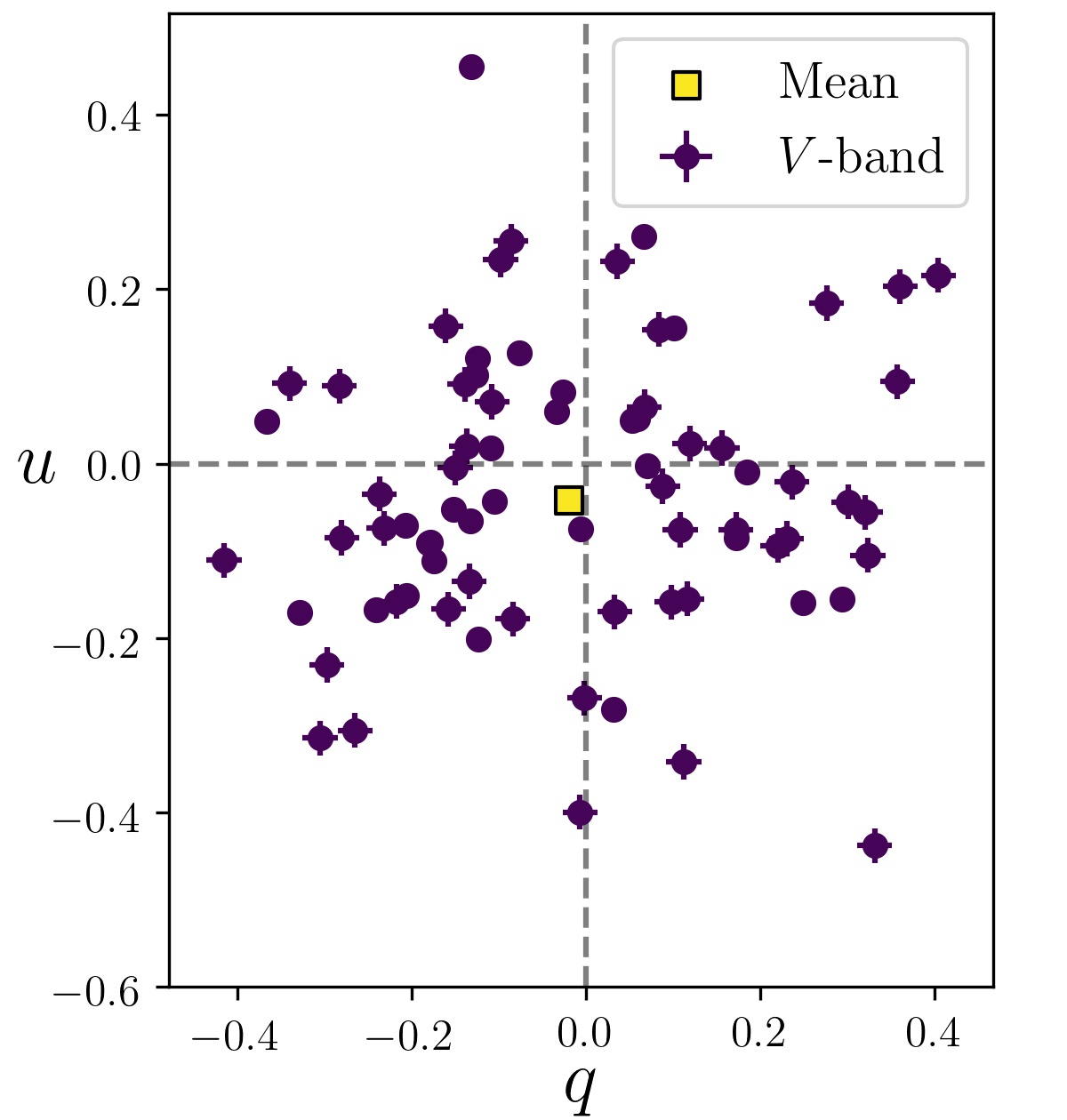}
    \caption{$q$ vs. $u$, for 76 intrinsic synthetic $V$-band observations of P Cygni after subtracting the ISP estimate described in \S\ref{sec:isp}. The yellow square marks the center of the distribution. The error of that mean location is smaller than the marker. While the separation between the mean of the $V$-band observations and the origin is statistically significant it is perhaps most notable how small this separation is. This indicates symmetry at the base of the wind in the P Cyg system---at least over the timescale of these observations.}
    \label{fig:iqu}
\end{figure}

\begin{figure}[!ht]
    \centering
    \plotone{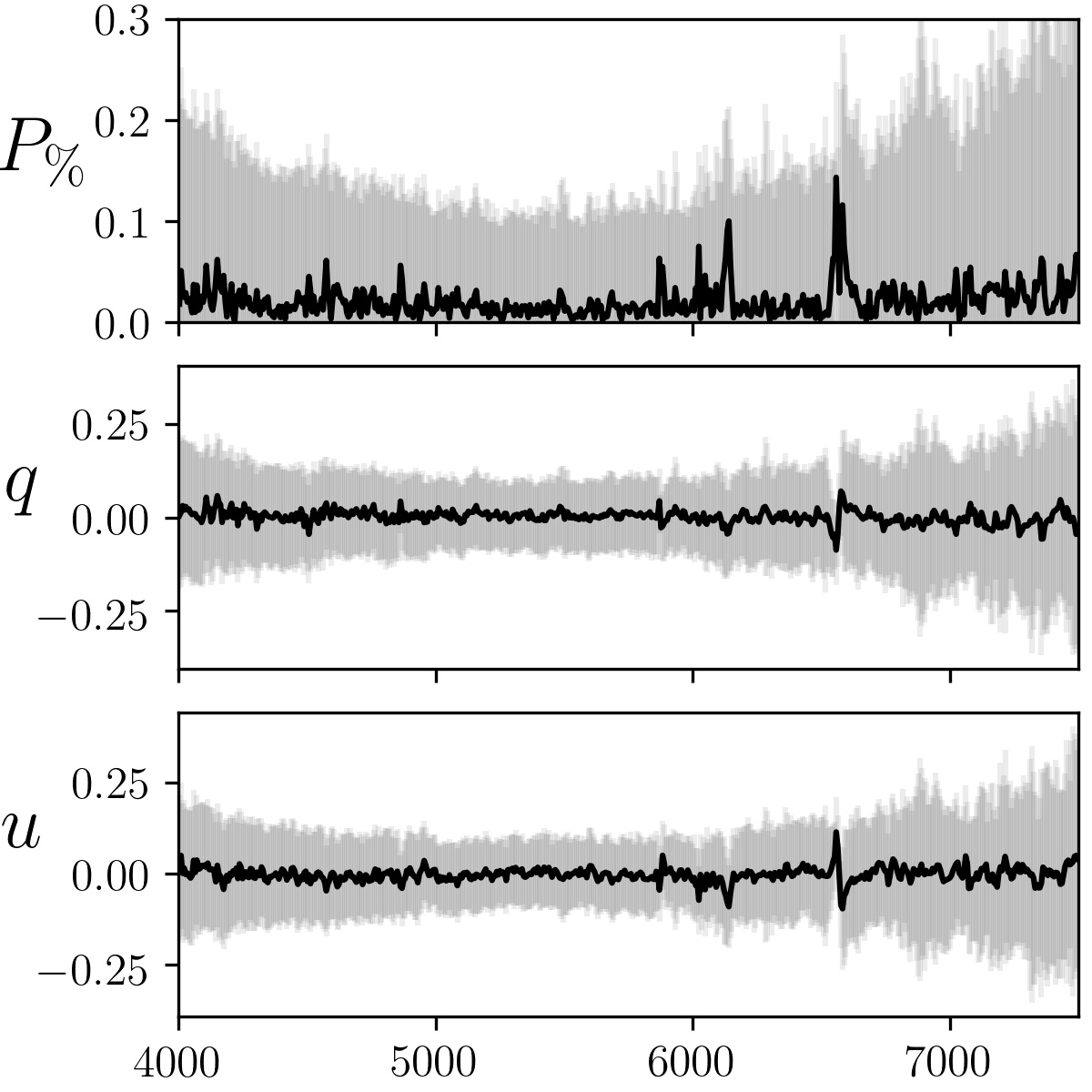}
    \caption{Mean intrinsic $P_\%$, $q$, and $u$ taken at each $\lambda$ bin over time. Means of percent polarization and position angle were calculated from the means of $q$ and $u$. The grey background indicates uncertainty of the mean measured at each $\lambda$ across all observations. $\Theta$ values are not included; since most $P_\%$ values are near the origin, position angle becomes highly uncertain.}
    \label{fig:intmean}
\end{figure}

\section{Results} \label{sec:results}

\begin{figure*}[ht!] 
    \plotone{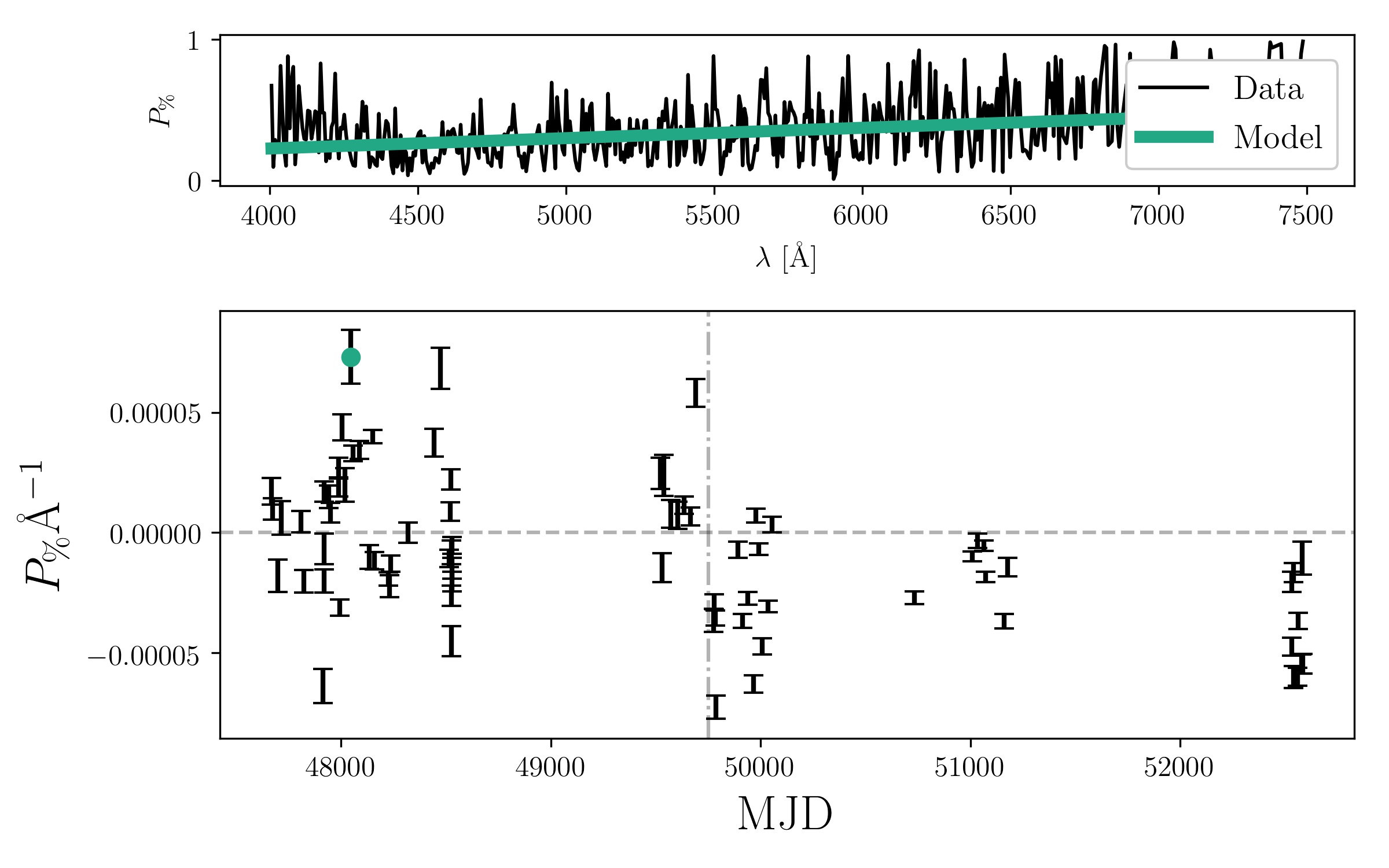}
    \caption{Slopes of lines fit to the $P_\%$ of each observation between 4000 and 7500 \AA{} plotted against time with 1$\sigma$ uncertainties. The dashed horizontal line indicates a slope of 0. The dashed vertical line shows the cutoff between Reticon observations (on the left) and CCD observations (on the right). The top panel shows the model (green) fit to the data (black) for the observation marked with a green dot.}
    \label{fig:slope}
\end{figure*}

\begin{figure}[!ht]
    \centering
    \plotone{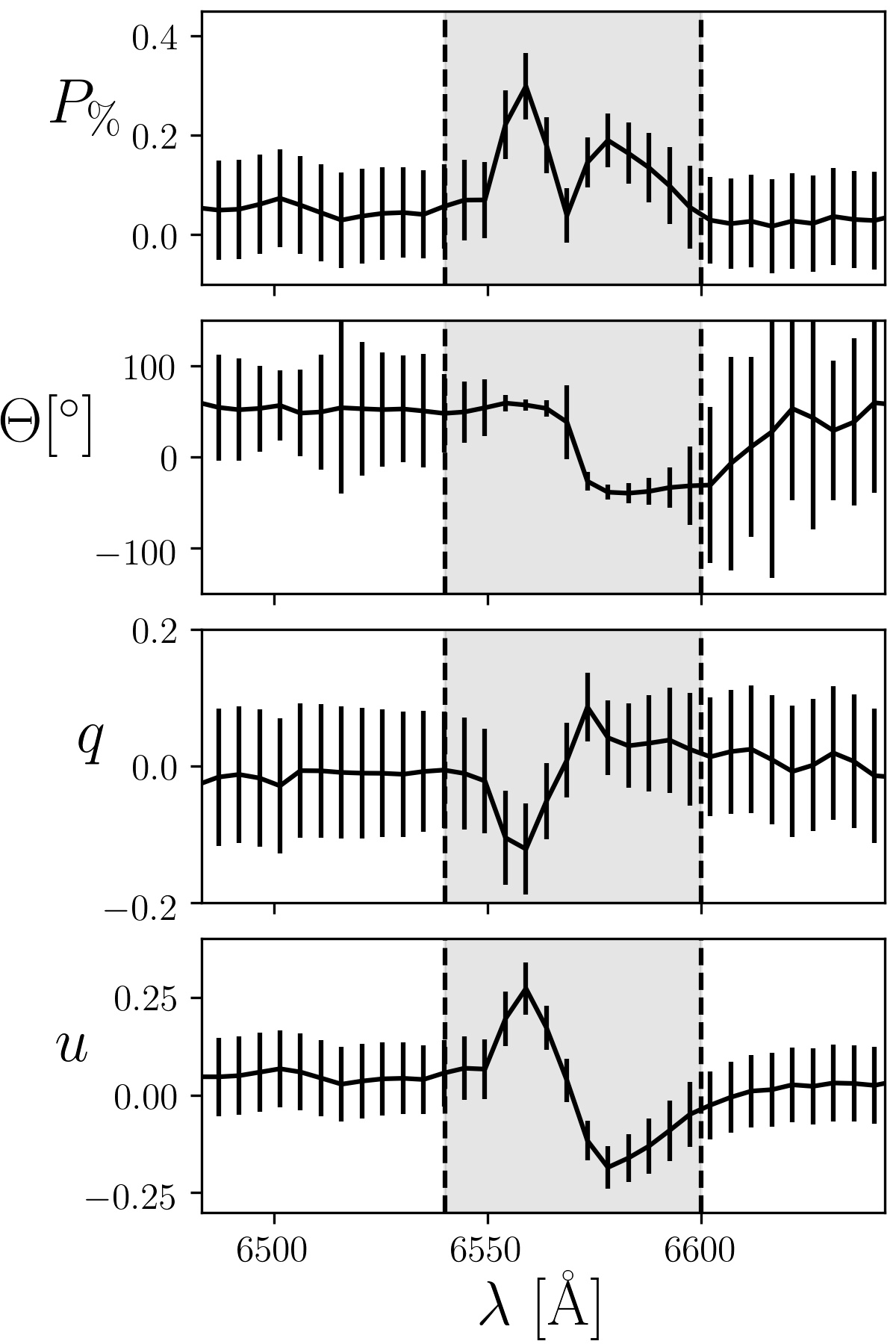}
    \caption{The mean, calculated over all CCD observations, of intrinsic $P_\%$, $\Theta$, Stokes $q$, and Stokes $u$ across H$\alpha$. The grey bounded region, between 6540 and 6600 \AA{} shows the line core from which the ISP at H$\alpha$ was calculated.}
    \label{fig:halpha}
\end{figure}

\begin{figure}[!ht]
    \centering
    \plotone{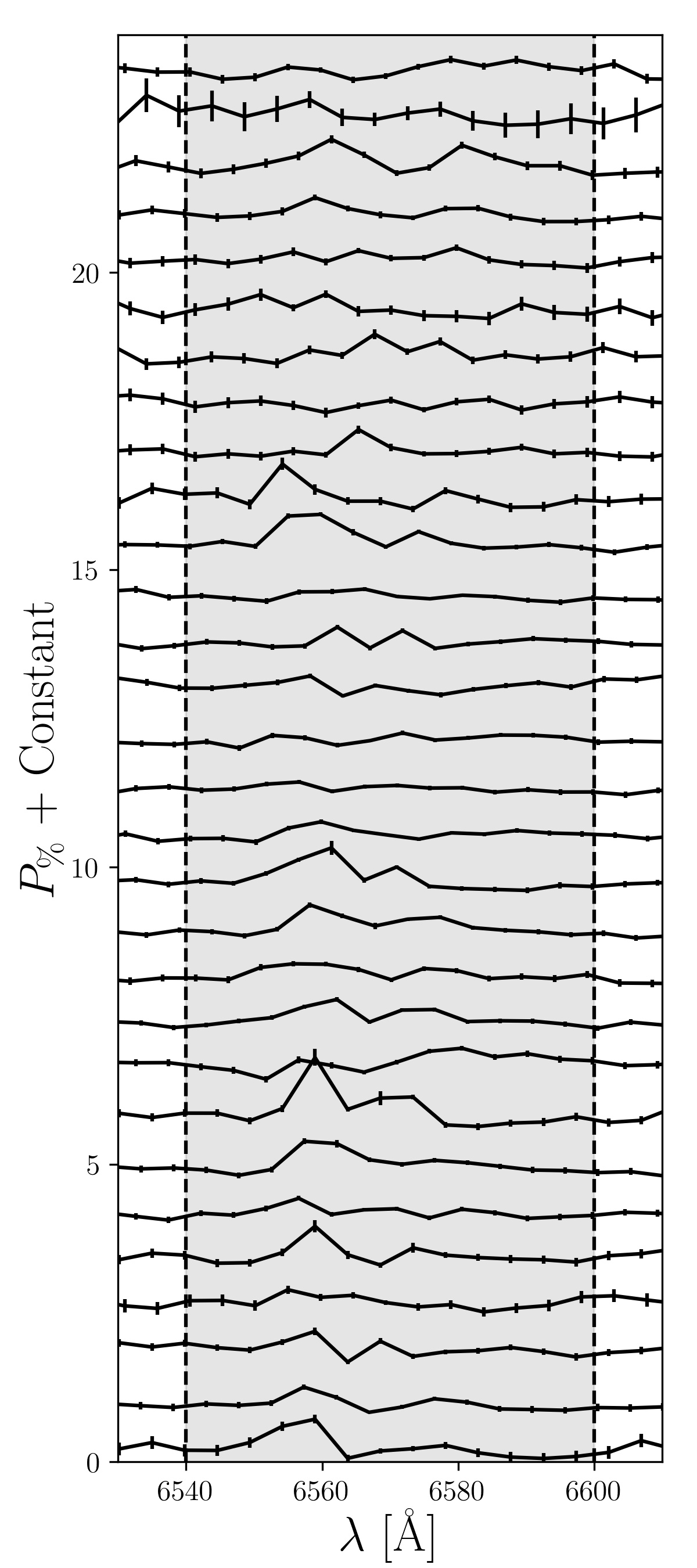}
    \caption{H$\alpha$ line-profile in percent polarization for 30 CCD observations. Each observation has a constant value added. The observations are stacked in chronological order, with the oldest being on the bottom. The grey bounded region, between 6540 and 6600 \AA{} shows the line core from which the ISP at H$\alpha$ was calculated. Note---the constant added does \textit{not} scale linearly with time.}
    \label{fig:stack}
\end{figure}

We used our ISP estimate determined above ($\lambda_{max}=$ 4712 \AA) and $P_{\%,ISP} = 1.06\%$) and equation \ref{eq:wrl} to calculate the wavelength dependence of the ISP. We converted that wavelength dependent $P_\%$ and $\Theta_{ISP}=34.4^\circ$ into wavelength dependent $q$ and $u$ values for the ISP. Subtracting those $q$ and $u$ components of the ISP from each observation results in an estimate of the intrinsic polarization of P Cygni. After we subtracted the ISP, we applied synthetic \textit{V}-band filters to our data (see \S\ref{sec:vband}) to quantify P Cyg's intrinsic continuum polarization behavior (Table \ref{HPOLObs}). After this subtraction, the center of the distribution of the \textit{V}-band data in $q$-$u$ space is shifted very near to the origin, as can be seen in Figure \ref{fig:iqu}. The distance between the center of that distribution and the origin is $0.047\pm0.002$. 

Additionally, we found the intrinsic polarization remaining in P Cyg's H$\alpha$ line and continuum (\S\ref{sec:pfew}; Table \ref{HPOLObs}). As discussed in \S\ref{sec:isp}, polarization in the H$\alpha$ line appears to be at least partially suppressed when compared to the \textit{V}-band, making it a good estimator of the ISP. However, we note that the H$\alpha$ line does show some variability with time, which suggests that not 100\% of its observed polarization is due to the ISP. Because we used the mean H$\alpha$ polarization to generate our ISP estimate, the possibility that we have over or under subtracted the ISP exists. Without a more reliable way to determine an ISP estimate we cannot evaluate how much of an effect this has on our intrinsic polarization of P Cyg, but the agreement of our $R_v$ calculated from our ISP estimate with $R_v$ values for P Cyg in the literature suggests our ISP estimate is robust and the temporal variability in the H$\alpha$ that remains after we performed ISP subtraction is due to changes in P Cyg's intrinsic polarization; i.e. on \textit{average} the polarization in P Cyg's H$\alpha$ line is a good estimator of the ISP, but night-to-night there can be a variable, intrinsic polarization component to the H$\alpha$ line. Therefore, we still report intrinsic $pfew$ H$\alpha$ line values in Table \ref{HPOLObs} for completeness and because the remaining signal in the lines likely holds key information about P Cyg's winds.

Figure \ref{fig:intmean} shows the mean $P_\%$, $q$, $u$, and $\Theta$ spectra, taken at each wavelength bin over time. Overall, these mean spectra are quite flat---as is expected for polarization caused by free electron scattering. We also see sharp features, most noteably at H$\alpha$. The magnitude of polarization is indistinguishable from zero unless averaged over a wide wavelength bin, as is the case for the $V$-band mean shown in Figure \ref{fig:iqu}. This shows that---at least when considered over a 13 year period---the ambient wind of P Cygni is nearly symmetrical on the plane of the sky. Note that uncertainties become significantly smaller at H$\alpha$, this is likely due to the increased number of photons received at the emission line. The region around $H\alpha$ is shown in more detail in Figure \ref{fig:halpha}.

\subsection{Intrinsic Polarization: the Search for Slope}

Both N01 and T91 describe a decrease of polarization levels moving into the infrared, which may be connected to the free-free absorptive opacity---$\kappa_{ff}$---in the wind of P Cygni. This observed slope is used in N01 to estimate the density of the inhomogeneities which are thought to give rise to intrinsic polarization in P Cygni. In order to investigate this previously discovered trend we fit a straight line to our data, this serves as a proxy for the true relation for the attenuation of polarization caused by $\kappa_{ff}$---$P_\% \propto e^{a\lambda^2}$. We perform this fit to our data between 4000 and 7500 \AA, after subtracting the ISP estimate found in \S\ref{sec:isp}, using \texttt{SciPy}'s \textit{curve}\_\textit{fit} function \citep{Jones2001SciPy:Python}. This function utilizes a least-squares fit of the form $y=mx+b$. One example of such a fit is shown in the upper panel of Figure \ref{fig:slope}. 

We report the results of this analysis in Figure \ref{fig:slope} as a slope ($P_\%$\AA$^{-1}$) for each observation with 1 $\sigma$ uncertainties. Because the exact value of the slope for each individual observation is highly dependent on our ISP calculation, it is most useful to look for trends across our observations. The slope of $P_\%$ does appear to change between individual observations at a statistically significant level. However, using \texttt{Astropy}'s \citep{Robitaille2013Astropy:Astronomy,Price-Whelan2018ThePackage} Lomb-Scargle analysis software\footnote{All Lomb-Scargle analysis was done using the default normalization as of \texttt{Astropy v. 4.0.1}} \citep{Lomb1976Least-squaresData,Scargle1982StudiesData,Press1989FastData} on these slopes showed no statistically significant periodicity, which is consistent with the stochastic variability displayed in Figure \ref{fig:slope}.

There do appear to be differences in the slopes of the $P_\%$ in our observations taken with HPOL's CCD  (to the right of the vertical dashed line in Figure \ref{fig:slope}) when compared to the Reticon detector (to the left of the vertical dashed line). The CCD observations have a more consistently negative slope, which corresponds to a persistent trend of a decrease in the $P_\%$ into the infrared consistent with N01's findings. However, the observations taken with HPOL's Reticon detector have a larger overall scatter in their slope, many of which are consistent with no wavelength dependence at the 3$\sigma$ uncertainty level. Some of our Reticon observations also show an increasing wavelength dependence into the infrared (positive slope). It is possible the overall polarization behavior of P Cygni changed when HPOL's detector was upgraded, no major changes in the system have been reported to be observed in other data sets corresponding to this time frame. Still, it cannot be ruled out that the difference in wavelength dependence between these two detectors is due to instrumental differences, perhaps due to differences in sensitivities between them.

\subsection{H$\alpha$ Polarization}
While it is true that intrinsic polarization at H$\alpha$ appears to be suppressed, as we show in Figure \ref{fig:comp}, there appear to be more complicated processes at work. Closer inspection of the H$\alpha$ line reveals a feature in the polarization spectrum of P Cygni. As shown in Figure \ref{fig:halpha}---though taking the mean washes out some of this feature---and Figure \ref{fig:stack}, there is clear structure at H$\alpha$ in both $P_\%$ and $\Theta$. While less easily seen in the Reticon data, this feature has an amplitude larger than the mean level of uncertainty of the CCD observations, suggesting that it is astrophysical in nature. The typical uncertainty of the Reticon observations is large enough that this feature is not clearly visible in many of those observations.

Figure \ref{fig:halpha} paints a puzzling picture. The strongest polarization appears near to the peak of the H$\alpha$ emission line. Moving red-ward there is a strong and rapid rotation, passing near to the origin in $q$ vs. $u$ space.

This feature shows temporal variations with no clear periodicity with regard to amplitude or shape (Figure \ref{fig:stack}). There are no significant correlations between the amplitude of the feature and the average percent polarization or position angle across the line. 

However, our Figure \ref{fig:slope} shows possible night-to-night changes in the wavelength dependence of P Cygni's polarization, which was also noted by T91 and N01, suggesting there are changes in P Cygni's free-free absorptive wind opacity. Because it is plausible that this absorption is also affecting the H$\alpha$ line and varying with time, we do not attempt to correct for it. Instead, we attempt to quantify variability in the line---which may be due to changes in its underlying absorption---using the methods described below.

\subsection{Periodicity} \label{sec:per}

Over the past century observers have noted irregular variability in P Cygni \citep{deGroot1969OnCygni}. As \citet{1936ZA.....11..304K} put it, ``P Cygni --- Nova of 1600 --- is one of the most remarkable stars of the Northern sky. It has attracted the attention of astronomers by an unusual character of light variation." More recently, however, evidence for periodicity in these unusual light variations has been presented by various authors \citep{deGroot2001CyclicitiesStars,deGroot2001CyclicitiesCygni,2002JAD.....8R...8V,2018NewA...65...29M,Kochiashvili2018OnCygni}. The periods found range from $\sim17$ day microvariations \citep{deGroot2001CyclicitiesCygni,deGroot2001CyclicitiesStars} to $\sim4.7$ years \citep{2018NewA...65...29M}. As will be discussed in \S\ref{dis:per}, there are many fascinating phenomenae that can cause periodic signals in polarimetric data. Here we detail our search for such signals.

\begin{figure*}
    \centering
    \plotone{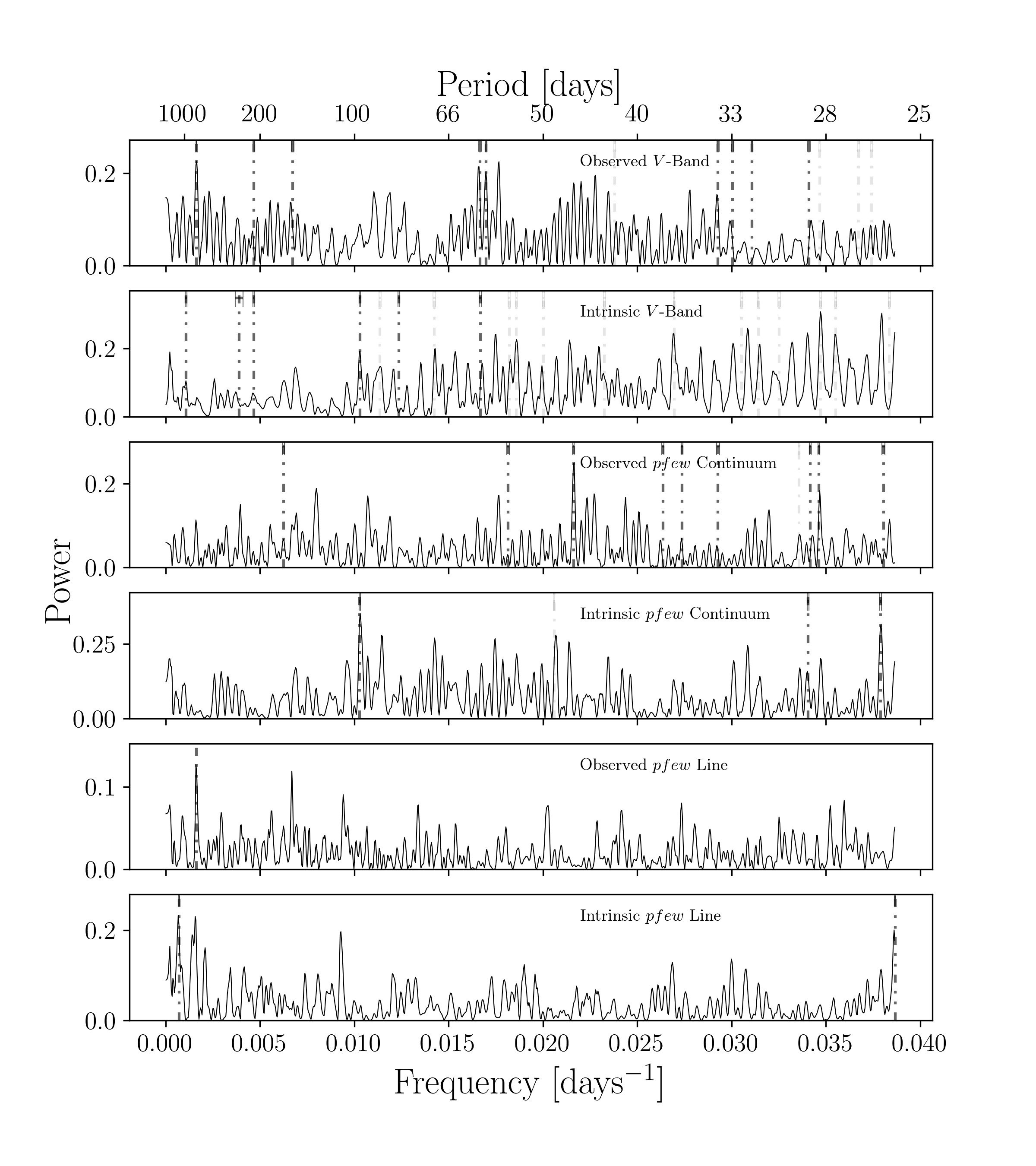}
    \caption{Lomb-Scargle periodograms of $P_\%$ data. Frequencies picked by the prewhitening procedure described in \S\ref{sec:per} are marked by vertical lines. Fainter lines indicate locations of frequencies which are harmonics or linear combinations of the darker, base frequencies (marked with $\dagger$ in Table \ref{tab:per}). Errorbars at the top of each vertical line indicate error in the frequency of the peak selected.}
    \label{fig:polper}
\end{figure*}

\begin{figure*}
    \centering
    \plotone{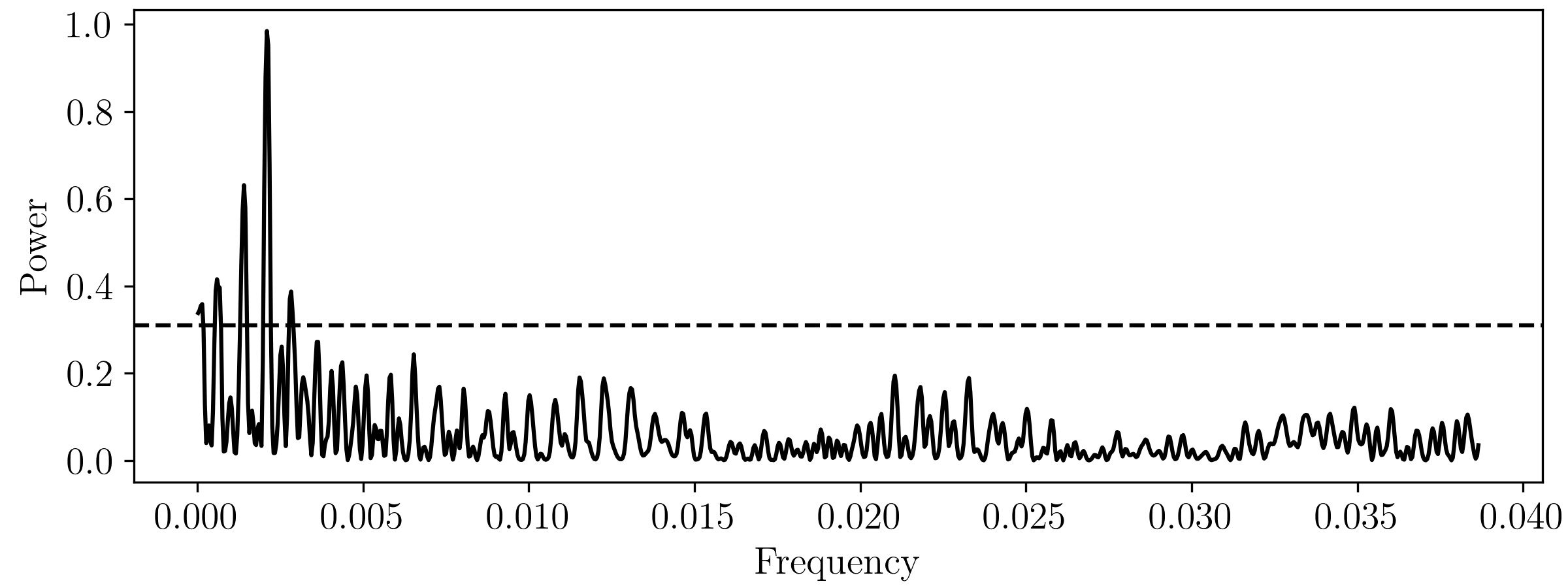}
    \caption{Lomb-Scargle periodogram analyzing simulated data which has the same temporal spacing of observations in Table \ref{HPOLObs}, based on a sinusoidal model with a period of 475 days. The horizontal dashed line shows the $\frac{1}{1,000}$ false alarm level.}
    \label{fig:testper}
\end{figure*}

We use \texttt{Astropy} \citep{Robitaille2013Astropy:Astronomy,Price-Whelan2018ThePackage} to calculate the Lomb-Scargle periodogram \citep{Lomb1976Least-squaresData,Scargle1982StudiesData,Press1989FastData} for all available sets of $P_\%$ data from Table \ref{HPOLObs}. Some sets of $\Theta$ data were found to have periodicities, these periods were significantly less robust under the prewhitening procedure described below.

We first search for peaks in the periodograms with low false alarm probability (FAP) under the null hypothesis of white noise \citep{Horne1986ASeries}. This results in multiple detected periods with FAP $\leq\frac{1}{1000}$. However, not all of these periodicities may be the result of astrophysical phenomena. As discussed extensively in \citet{VanderPlas2018UnderstandingPeriodogram}, there are many subtleties in interpreting the Lomb-Scargle periodogram. Therefore, we also conduct a prewhitening procedure, similar to that described in \citet{Dorn_Wallenstein_2019}. In short, we iteratively select the highest peak in the periodogram, fit a sinusoid to the data (allowing the frequency to vary within the resolution of the periodogram), subtract the sinusoid, and recalculate the periodogram until we reach a minimum in the Bayesian Information Content of the fit \citep{Schwarz1978EstimatingModel}. Formal errors on the frequency, amplitude, and phase of the sinusoids extracted in each stage of prewhitening are calculated following \citet{1971AJ.....76..544L} and \citet{1999DSSN...13...28M}. This results in a list of frequencies that uniquely describe the data to within the noise. Frequencies extracted via prewhitening are listed---in the order that they were detected by the prewhitening algorithm---in Table \ref{tab:per}, and the full periodograms are shown in Figure \ref{fig:polper}. We additionally search for harmonic frequencies---$f_i = nf_\circ$ where n is an integer---and combination frequencies of the form $f_i = f_1 + f_2$.

These periodograms have relatively poor signal-to-noise ratios (SNRs), and many of the extracted frequencies have large false alarm probabilities, meaning that it is difficult to say whether any particular period is ``real" or simply a statistical artefact. However, several periods are more convincing than others. For example, the 97 day period found in the $pfew$ continuum $P_\%$ data has a small FAP (of order $10^{-4}$), is the base of multiple harmonic/combination frequencies, and is also detected in the intrinsic $V$-band $P_\%$ data.

\begin{center}
\startlongtable
\begin{deluxetable*}{c|lllll} 
\label{tab:per}
\tablehead{\colhead{\textbf{Data}} &   
\colhead{\textbf{Frequency} [days$^{-1}$]} & 
\colhead{\textbf{Frequency Error}} & 
\colhead{\textbf{Period} [days]} &  
\colhead{\textbf{SNR}} &  
\colhead{\textbf{False Alarm Probability}}}
\startdata
\multirow{13}{*}{Observed $V$-Band}
& 0.001629$^\dagger$ & 0.000031 & 614.0 & 5.0 & 0.048 \\ 
& 0.016990$^\dagger$ & 0.000038 & 58.9 & 5.6 & 0.129 \\ 
& 0.030053$^\dagger$ & 0.000055 & 33.3 & 3.7 & $\sim$1 \\ 
& 0.037419 & 0.000041 & 26.7 & 5.1 & $\sim$1 \\ 
& 0.036741 & 0.000061 & 27.2 & 4.9 & $\sim$1 \\ 
& 0.023797 & 0.000045 & 42.0 & 5.7 & $\sim$1 \\ 
& 0.004683$^\dagger$ & 0.000045 & 213.5 & 5.3 & $\sim$1 \\ 
& 0.034663 & 0.000040 & 28.8 & 4.6 & $\sim$1 \\ 
& 0.006739$^\dagger$ & 0.000067 & 148.4 & 4.9 & 0.874 \\ 
& 0.031080$^\dagger$ & 0.000055 & 32.2 & 9.4 & $\sim$1 \\ 
& 0.029281$^\dagger$ & 0.000064 & 34.2 & 5.4 & 0.647 \\ 
& 0.016665$^\dagger$ & 0.000074 & 60.0 & 5.5 & 0.083 \\ 
& 0.034099$^\dagger$ & 0.000071 & 29.3 & 5.0 & $\sim$1 \\ 
\midrule
\multirow{19}{*}{Intrinsic $V$-Band}
& 0.034717 & 0.000035 & 28.8 & 4.6 & 0.001 \\ 
& 0.026951 & 0.000032 & 37.1 & 7.0 & 0.023 \\ 
& 0.018596 & 0.000036 & 53.8 & 4.1 & 0.049 \\ 
& 0.010299$^\dagger$ & 0.000038 & 97.1 & 6.9 & 0.206 \\ 
& 0.020022 & 0.000049 & 49.9 & 4.7 & 0.927 \\ 
& 0.031412 & 0.000042 & 31.8 & 5.1 & 0.117 \\ 
& 0.035521 & 0.000047 & 28.2 & 6.0 & 0.025 \\ 
& 0.032523 & 0.000069 & 30.7 & 5.5 & $\sim$1 \\ 
& 0.011359 & 0.000055 & 88.0 & 4.7 & 0.744 \\ 
& 0.038363 & 0.000061 & 26.1 & 4.1 & $\sim$1 \\ 
& 0.030532 & 0.000058 & 32.8 & 7.4 & $\sim$1 \\ 
& 0.012354$^\dagger$ & 0.000045 & 80.9 & 6.4 & $\sim$1 \\ 
& 0.004669$^\dagger$ & 0.000039 & 214.2 & 4.9 & $\sim$1 \\ 
& 0.001084$^\dagger$ & 0.000044 & 922.9 & 5.6 & $\sim$1 \\ 
& 0.016677$^\dagger$ & 0.000059 & 60.0 & 5.0 & 0.811 \\ 
& 0.018210 & 0.000061 & 54.9 & 6.1 & 0.455 \\ 
& 0.014242 & 0.000066 & 70.2 & 5.5 & 0.152 \\ 
& 0.023261 & 0.000058 & 43.0 & 4.1 & 0.948 \\ 
& 0.003896$^\dagger$ & 0.000200 & 256.6 & 5.2 & $\sim$1 \\ 
\midrule
\multirow{10}{*}{Observed $pfew$ Continuum}
& 0.021624$^\dagger$ & 0.000037 & 46.2 & 5.1 & 0.018 \\ 
& 0.034627$^\dagger$ & 0.000038 & 28.9 & 5.4 & 0.280 \\ 
& 0.027373$^\dagger$ & 0.000038 & 36.5 & 6.4 & $\sim$1 \\ 
& 0.034168$^\dagger$ & 0.000057 & 29.3 & 5.2 & $\sim$1 \\ 
& 0.006258$^\dagger$ & 0.000044 & 159.8 & 6.3 & $\sim$1 \\ 
& 0.033576 & 0.000052 & 29.8 & 8.4 & $\sim$1 \\ 
& 0.029280$^\dagger$ & 0.000083 & 34.2 & 4.6 & $\sim$1 \\ 
& 0.038049$^\dagger$ & 0.000085 & 26.3 & 4.4 & $\sim$1 \\ 
& 0.026358$^\dagger$ & 0.000053 & 37.9 & 6.0 & $\sim$1 \\ 
& 0.018158$^\dagger$ & 0.000064 & 55.1 & 7.1 & $\sim$1 \\ 
\midrule
\multirow{4}{*}{Intrinsic $pfew$ Continuum}
& 0.010285$^\dagger$ & 0.000035 & 97.2 & 5.0 & 0.0001 \\ 
& 0.037901$^\dagger$ & 0.000045 & 26.4 & 5.5 & 0.0007 \\ 
& 0.020591 & 0.000033 & 48.6 & 5.0 & 0.012 \\ 
& 0.034052$^\dagger$ & 0.000053 & 29.4 & 4.3 & 0.591 \\ 
\midrule
\multirow{1}{*}{Observed $pfew$ Line}
& 0.001637$^\dagger$ & 0.000041 & 611.0 & 6.2 & 0.965 \\ 
\midrule
\multirow{2}{*}{Intrinsic $pfew$ Line}
& 0.000720$^\dagger$ & 0.000025 & 1388.2 & 3.3 & 0.118 \\ 
& 0.038670$^\dagger$ & 0.000039 & 25.9 & 7.7 & 0.162 \\ 
\enddata
\tablecomments{$^\dagger$ mark base frequencies---which are $not$ a harmonic or linear combination of other frequencies.}
\end{deluxetable*}
\end{center}

We also note that several of the base periods (those marked with a $\dagger$ in Table \ref{tab:per}) have multiple harmonics of the same order. This could be a result of errors in either the base or harmonic frequencies. In fact, the presence of any harmonics may be evidence that any true periodic oscillations are not sinusoidal. If the polarization in P Cygni arises from the ejection of asymmetries into the ambient stellar wind, as is described in T91 and N01, we should not expect the time-series $P_\%$ curve to be a sinusoid. Under these conditions that curve would peak sharply when such an asymmetry is ejected from the photosphere.  

\subsubsection{Cadence Test}
To determine if this observed periodicity is a result of the temporal spacing of the observations, we perform a Lomb-Scargle periodicity search on simulated data which lacks any of the periods that we have found. We claim that if we detect a similar period in this test, then the periods found in Table \ref{tab:per} can be attributed to how the data was sampled, not an astrophysical process.

To create a set of simulated data we used a series of periodic functions with various periods (for example $\sin(\frac{2\pi t}{475})$), as well as flat distributions. We calculated the values of these functions at the time values of our observations (MJD) then added Gaussian noise defined by errors at each observation in Table \ref{HPOLObs}.

However, as is exemplified in Figure \ref{fig:testper}, all incorrect periods found in such data sets are either clustered around the true period or a harmonic of the true period. Based on this result, we conclude that the periodicities must be either astrophysical in nature or a result of more complicated instrumental/analytical effects. The likelihood of our findings being due to HPOL is minimal as HPOL is a well characterized instrument and spurious periodicities have not been found in other HPOL data sets. But, as can be seen in Table \ref{tab:per}, there are different periods found in intrinsic polarization as compared to observed polarization. This could be a sign that the process of subtracting out the interstellar polarization signal (see \S\ref{sec:isp}) may either obscure or insert periodicities. For example, in $q$-$u$ space, the $P_\%$ of a circle centered at the origin would be constant, whereas one shifted away from the origin would result in a sine wave pattern in $P_\%$.

Additionally, converting the MJD values in Table \ref{HPOLObs} to \textit{barycentric} Julian date had no effect on any of the results discussed here.

\subsubsection{pfew Periodicity}
By searching for periodicity in \textit{pfew} H$\alpha$ \textit{line} data, we can test whether there is any non-stochastic variability present in P Cyg's H$\alpha$ emission line itself, as opposed to the \textit{V} band continuum or the \textit{pfew continuum}. Figure \ref{fig:polper} shows our resulting Lomb-Scargle periodograms. The evidence for any significant periods is much weaker than in the continuum data. This suggests that any periodicities are not due to changes in the H$\alpha$ emission region of P Cygni.

\subsection{Ellipticity in $q$-$u$ Space: The Mauchly Test} \label{sec:mly}
The position angle of P Cygni’s intrinsic polarization appears to move randomly, with no strong preference for one orientation. In other words, whatever mechanism is responsible for P Cygni’s mass loss appears to eject mass asymmetries stochastically in all directions. This can be seen from a $q$-$u$ plot. A random distribution of position angles will produce a circular distribution of points in q vs. u space, as appears to be the case in Figures \ref{fig:comp} and \ref{fig:iqu}.

Here, we attempt to quantify the statistical significance of ellipticity of the \textit{V} band \textit{q-u} distribution using the Mauchly test \citep{Mauchly1940SignificanceDistribution} to determine if their position angle is truly random (i.e. circular in \textit{q-u} space) or if there is a slightly preferred position angle (i.e. elliptical in \textit{q-u} space). The Mauchly test has been used extensively in the field of biostatistics \citep{Davis2003StatisticalMeasurements,Myers2010ResearchAnalysis,Crowder2017AnalysisMeasures}, but has thus far not been used in the context of polarimetry.
 
 In the original paper by Mauchly, he seeks to identify significant ellipticity in a collection of two-dimensional data points. A collection of data points in two dimensions may appear elliptically distributed even when their parent population is actually circular, simply due to random sampling. Thus, Mauchly asks ``What is the probability of obtaining, from a circular normal population, a random sample of N points for which the ellipticity is as great or greater than that actually obtained in the given sample of N points?'' \citep{Mauchly1940ADial}.
 
To help answer this question, Mauchly defines the ellipticity statistic:
\begin{equation}
\mathcal{L}_e= \frac{2\sigma_q\sigma_u\sqrt{1-r^2}}{\sigma_q^2 + \sigma_u^2},
\label{eq:le}
\end{equation}
where $r$ is Pearson's correlation coefficient between the distribution in $q$ and $u$, and where $\sigma^2_q$ and $\sigma^2_u$ are the sample variances. By using $r$, $\sigma^2_q$, and $\sigma^2_u$, this test assumes that the distribution can be described by a two dimensional Gaussian, defined by the covariance matrix, 
\[
K_{qu} = 
\begin{pmatrix}
    \sigma^2_q & r \\
    r & \sigma^2_u
\end{pmatrix}.
\]
Note that if the distribution of points is circular, then $\sigma^2_q = \sigma^2_u$, $r=0$, and $\mathcal{L}_e=1$. Conversely, if the distribution of points is elliptical, then $\sigma^2_q \neq \sigma^2_u$, $r\rightarrow 1$, and $\mathcal{L}_e\rightarrow 0$.

Assuming that $N$ points are drawn from a circular distribution, the statistic $\mathcal{L}_e$ has the distribution
\begin{equation}
    f(\mathcal{L}_e) = (N-2)\mathcal{L}_e^{N-3} d\mathcal{L}_e.
\end{equation}
If the actual value of $\mathcal{L}_e$ calculated from the data is $\mathcal{L}_e$, then the probability that a value as small or smaller than $\mathcal{L}_e$ is found from a random sample of $N$ points is
\begin{equation}
    P( \mathcal{L}_e ) = \int_0^{\mathcal{L}_e} (N-2)\mathcal{L}_{e}^{N-3}d\mathcal{L}_e= \mathcal{L}_e^{N-2}.
\end{equation}
When $P( \mathcal{L}_e )$ is close to one, then the null hypothesis cannot be rejected (i.e. the parent distribution is likely circular) \citep{Mauchly1940ADial}. We take the position that a value of $P( \mathcal{L}_e )$ which is less than $\alpha = 0.05$ provides sufficient evidence to reject the null hypothesis.

To illustrate how these numbers are interpreted, Table \ref{tab:ellipticity} includes three general examples in the last three rows. The first is a distribution which is very close to being circular, however, the small amount of ellipticity present is statistically significant. The next is an extremely elliptical distribution which would appear to be a line. But, there is a 50\% chance that points drawn from a circular distribution would appear more elliptical. The last example is an elliptical distribution which is very statistically significant; this example would be similar to what we would see for a significant preferred position angle.
\begin{figure}[ht!]
    \centering
    \plotone{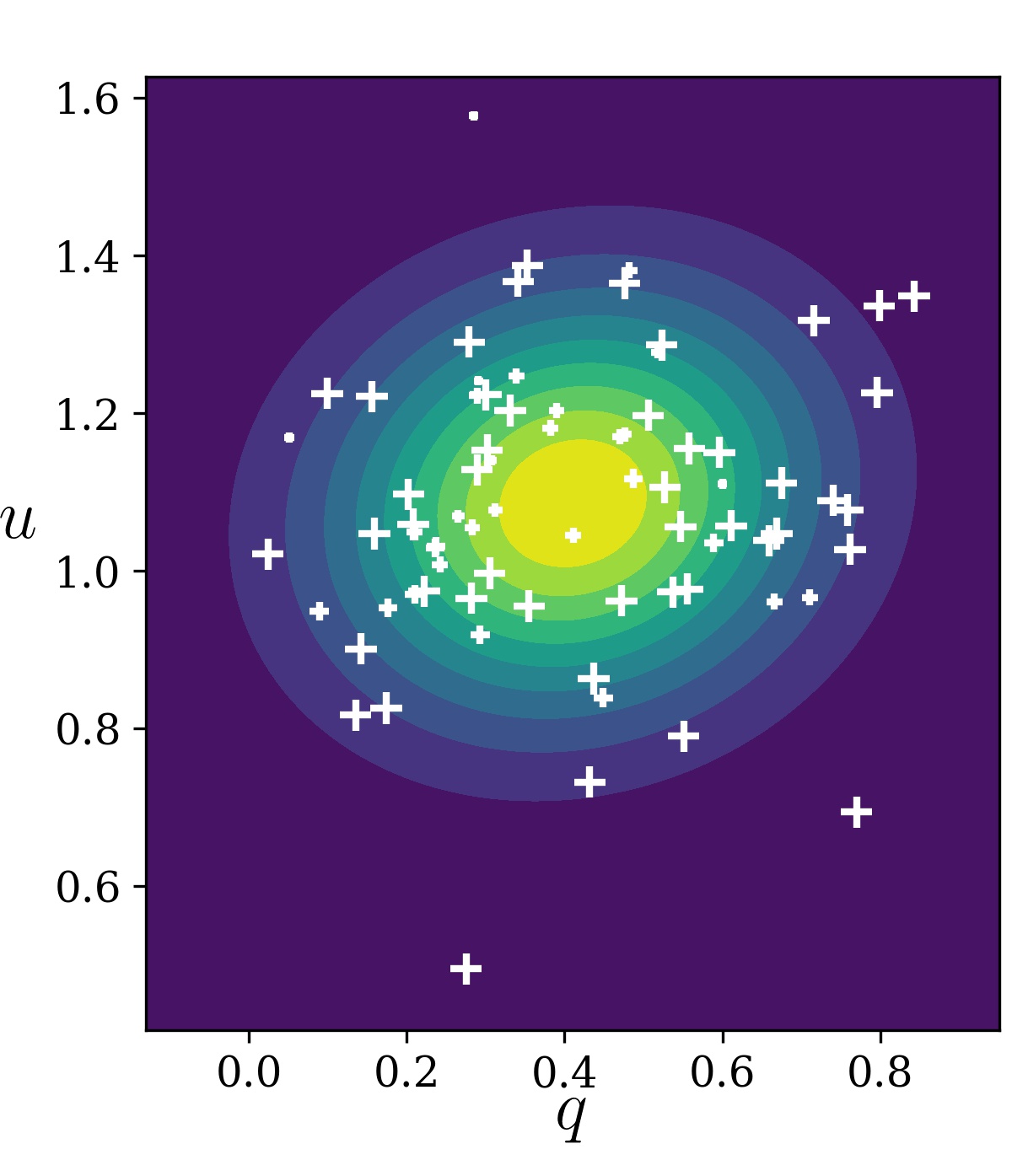}
    \caption{White crosses show the position of observed $V$-band observations of P Cygni in $q$ vs. $u$ space, with the size of the cross indicating uncertainty. Underlying colormap shows the 2D Gaussian distribution derived from $r$, $\sigma_q$, and $\sigma_u$ (discussed in \S\ref{sec:mly}). Contours show equally spaced slices in likelihood, the lightest colors corresponding to a peak in the probability distribution function.}
    \label{fig:gauss}
\end{figure}

Using all $q$-$u$ data sets from Table \ref{HPOLObs} we apply this Mauchly test. For the observed $V$-band values, plotted in Figure \ref{fig:gauss}, the sample standard deviations and Pearson correlation coefficient are $\sigma_q = 0.202 ;~\sigma_u = 0.177 ;~r = 0.115$. Given these values we find that $\mathcal{L}_e = 0.985$ and $P(\mathcal{L}_e) = 0.320$. This is not significant evidence for a preferred position angle and is consistent with the findings of T91 and N01. In fact, analyzing any other data set in Table \ref{HPOLObs} in this manner yields similar results. The fact that the observed $pfew$-line observations do not show evidence for a preferred position angle actually strengthens the argument for using it as an ISP probe, since deviations will then be evenly distributed about the desired value.

\subsubsection{A Test of the Mauchly Test} \label{sec:mtest}
To see if the Mauchly Test is capable of finding a preferred position angle we use the method described above on a system with a known preferred position angle. Here we use \textit{HPOL} data from \citet{Hoffman1998,Lomax2012GEOMETRICALLYRAE} of $\beta$ Lyrae---a self-eclipsing binary with mass transfer via an accretion disk. The presence of this disk creates net linear polarization with a preferred position angle perpendicular to the disk.

\begin{figure*}[ht!]
    \centering
    \plotone{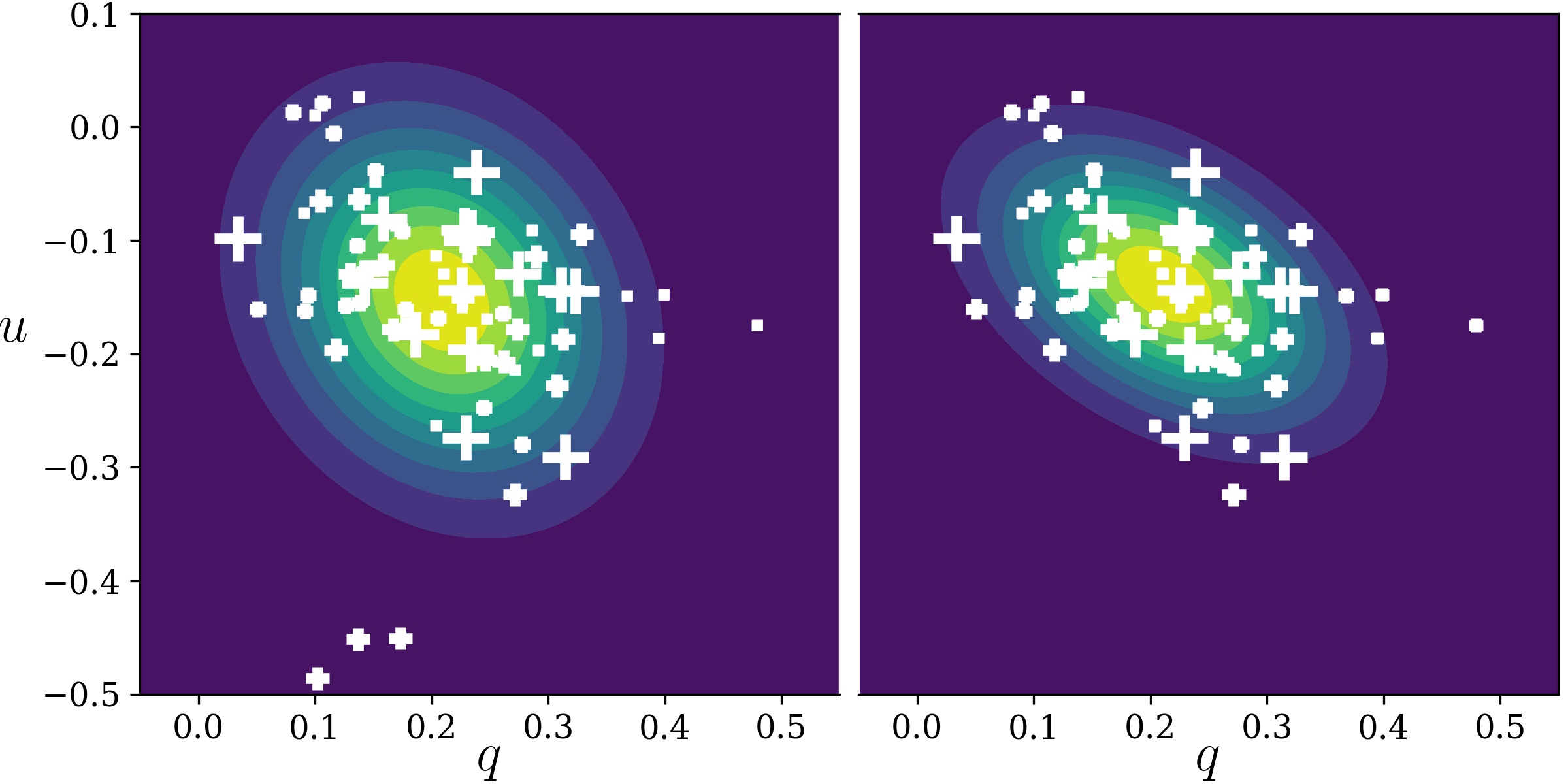}
    \caption{Stokes $q$-$u$ data for a test system with a preferred position angle, $\beta$ Lyr from \citet{Lomax2012GEOMETRICALLYRAE}, see \S\ref{sec:mtest}. White crosses represent individual $V$-band observations, with the size of the cross indicating uncertainty, while the underlying colormap shows the distribution derived from $r$, $\sigma_q$, and $\sigma_u$ (discussed in \S\ref{sec:mly}). Contours of the color map are equally spaced slices in likelihood, the lightest colors corresponding to a peak in the probability distribution function. The left panel shows the distribution with the three outliers included, the right panel shows the distribution without those observations.}
    \label{fig:tgaussnc}
\end{figure*}

Using 69 non-ISP removed, $V$-band polarimetric observations from \citet{Lomax2012GEOMETRICALLYRAE}, we find the distribution shown in the left panel of Figure \ref{fig:tgaussnc}, with $\sigma_q = 0.089 ;~\sigma_U = 0.099 ;~r = -0.203$. This corresponds to $\mathcal{L}_e = 0.975$ and $P(\mathcal{L}_e) = 0.178$. This would seem to suggest that the Mauchly test would not be capable of detecting the present preferred position angle. 

However, there is obviously a set of outliers towards the bottom of the figure. These points are identified as outliers by \citet{Lomax2012GEOMETRICALLYRAE}, and all three observations were taken on the same night. When these outliers are discounted, the model of this distribution changes to that of the right panel of Figure \ref{fig:tgaussnc}. Using this subset of data, we derive the values $\sigma_q = 0.089 ;~\sigma_u = 0.074 ;~r = -0.443$. This corresponds to $\mathcal{L}_e = 0.880$ and $P(\mathcal{L}_e) = 0.0003$. which is significantly below our threshold of $\alpha=0.05$. Thus, we can reject the null hypothesis that the distribution is circular.

This example outlines an important limitation of this test. In small data-sets, the presence of even a small number of outliers can have an impact on the estimated value of $\mathcal{L}_e$ and the results of the associated significance test.

\begin{center}
\begin{deluxetable}{lll} 
\tablecolumns{3}
\tablewidth{0pt}
\tablecaption{Ellipticity \label{tab:ellipticity}}
\tablehead{\colhead{Data}& \colhead{$\mathcal{L}_e$} & \colhead{$P( \mathcal{L}_e )$}}
\startdata
P Cygni                  & 0.986 &    0.354 \\
$\beta$ Lyrae            & 0.975 &    0.178 \\
$\beta$ Lyrae (adjusted) & 0.880 & 0.0003 \\
\hline
Significant Circle     &        0.95 &       1e-05 \\
Insignificant Line  &        0.01 &         0.5 \\
Significant Ellipse       &      0.5 &       1e-05 \\
\enddata
\end{deluxetable}
\end{center}

\section{Discussion} \label{sec:dis}
These results allow us to infer properties of the material in the near circumstellar environment around P Cygni. Reporting on 13 years of observations of P Cygni we can revisit the findings of previous spectropolarimetric studies.

This paper supports previous claims from  T91, N01, and \citet{Hayes1985VariableCygni} that asymmetries in the system lack a preferred polarimetric position angle. Not only is the distribution of observations in $q$-$u$ space nearly circular, but the small amount of observed ellipticity is statistically insignificant as shown in Table \ref{tab:ellipticity}. Previous studies investigating the resolved nebula around P Cygni \citep{Leitherer1987TheImaging.,Barlow1994TheNebula,Nota1995NebulaePicture,Meaburn1996TheCygni} have reported that this nebula is roughly spherical in shape, with some clumping. This is not entirely expected as more than 50\% of resolved galactic LBVs have bipolar nebulae \citep{Weis2012LBVInstabilities}. Our findings suggest that this lack of observed directionality in the circumstellar environment starts at the base of the wind. This means that this mode of mass loss is either truly lacking any directional preference, or that the preferred direction is hidden by projection effects. It is possible, for example, that the mass inhomogeneities could be ejected solely from the equator of the star, but if the equatorial plane is close to the plane of the sky then it would be nearly impossible to tell.

It is more difficult to interpret the wavelength dependent feature observed at H$\alpha$. This is troubling, particularly in light of the importance of using the line center method for ISP estimates of early-type stars. The higher levels of polarization in the line (shown in Figure \ref{fig:halpha}) could potentially skew measurements of the ISP in studies of P Cygni and other strong emission-line objects. While this feature has a relatively small amplitude, and it likely does not have a significant impact on the analysis in this work, we cannot be certain as to the extent of its impact until it is explained.

\subsection{Problematic Polarimetric Periodicities} \label{dis:per}
The series of periodicities in the polarimetric data poses a serious question. What could cause these periodicities? While it is unlikely that all instrumental effects can be ruled out, HPOL was a well-characterized instrument. Therefore, for the remainder of this section we shall attempt to explain this result as a truly astrophysical phenomenon. 

We did not find strong evidence for periodicity in the H$\alpha$ emission line (using $pfew$ as described in \S\ref{sec:pfew}) or in the slopes of the individual observations' $P_\%$ versus wavelength data. This suggests that the cause of these periods is likely not due to changes in the free-free absorptive opacity of P Cygni's wind, which is thought to cause a general decrease in its $P_\%$ into the infrared. 

These results could be explored with great depth, and likely deserves their own paper, but here we will discuss the major points, and possible explanations for the periodic effects observed in these data. In particular, we focus the majority of this discussion on the 97 day period in the continuum polarization found in \S\ref{sec:per}, as that is the period with the strongest false alarm probability and has proven quite robust over different iterations of analysis.

\subsubsection{P Cygni Absorption}
It is interesting to note that there are periodicities which appear in the \textit{pfew} Continuum $P_\%$ that are not present in the $V$-Band $P_\%$ data. Given how close these two regions of continuum are, this is somewhat unexpected. If these differences are astrophysical in nature it points to a difference not in the H$\alpha$ emission line, but to a difference in what is happening at H$\alpha$ in continuum emission. 

A possible explanation could be the eponymous P Cygni absorption component of the H$\alpha$ line. Since our implementation of the \textit{pfew} method does not account for this absorption component, it should have some impact on the continuum values tabulated in Table \ref{HPOLObs}. Therefore, it is possible that we are detecting periodicity in the polarization of the absorption component itself.

\subsubsection{Pulsations}
The 97 day period listed in Table \ref{tab:per} lines up with $\sim100$-d type periods found by \citet{deGroot2001CyclicitiesStars} in a study of the photometric behavior of P Cygni over multiple decades. The authors attempt to explain this mode of variability as being a result of radial pulsations. They find an anti-correlation between $m_V$ and $B-V$; the star becomes redder as it gets brighter, as would be expected for radial pulsations where the increased brightness is the result of an increase in radius and subsequent drop in effective temperature. While it is not likely that radial pulsations would directly affect the net polarization that we observe, inhomogeneities in pulsation-driven mass loss from the star (expected when the star is in its cool state, with a more extended atmosphere and lower surface gravity) could lead to periodic variations in polarization.

Hydrodynamic models of LBV pulsations by \citet{Jiang2018OutburstsOpacity} have also predicted pulsations with periods of order $\sim$day timescales. While it is less likely that the data presented here would be sensitive to such short periods, these fast pulsations may combine with other effects to produce the periodicity which we see.

\subsubsection{Rotation}
It is possible that inhomogeneities are being sourced from a stationary feature on the photosphere, such as a hotspot, corotating interaction region, or some other localized site of asymmetric mass loss. Using a radius of 76$R_\sun$ and $v\sin{i} = 35$ kms$^{-1}$ \citep{Najarro1997ALines}, the rotational period of P Cygni would be roughly 110 days. Since free-electron scattering produces maximum polarization for a 90$^\circ$ scattering angle, we would observe the effect of asymmetries most strongly when this feature is at the plane of the sky.

However, we would expect to see peaks in polarization twice per rotation period for this scenario to make sense in the context of our data set---which is a higher frequency than most of the base frequencies we have found. Additionally, it is not certain that a feature such as a hotspot or mass loss asymmetry can be stable over a 13 year period.

\subsubsection{A Companion} \label{dis:com}
The potential binarity of LBVs is a contentious issue, but one explanation for the period found in this data is that P Cygni has a binary companion. Periodic variability in polarimetric data arising from companions is not new. For example, Wolf-Rayet-O-star and Roche-lobe overflow binary systems show variability in this manner (see \citealt{Brown1978PolarisationInclinations,Moffat1993PolarizationBinaries,Hoffman1998,Lomax2014}).

Assuming the masses of P Cygni and its possible companion are each between 20 and 60 $M_\sun$, we can place some rough limits on the geometry of the system. For these masses, and a 97 day orbital period, we find a semi-major axis of roughly 1-2 AU, or about 3-6 times the radius of P Cygni \citep{Najarro2001SpectroscopyCygni}.

\begin{figure}
    \centering
    \plotone{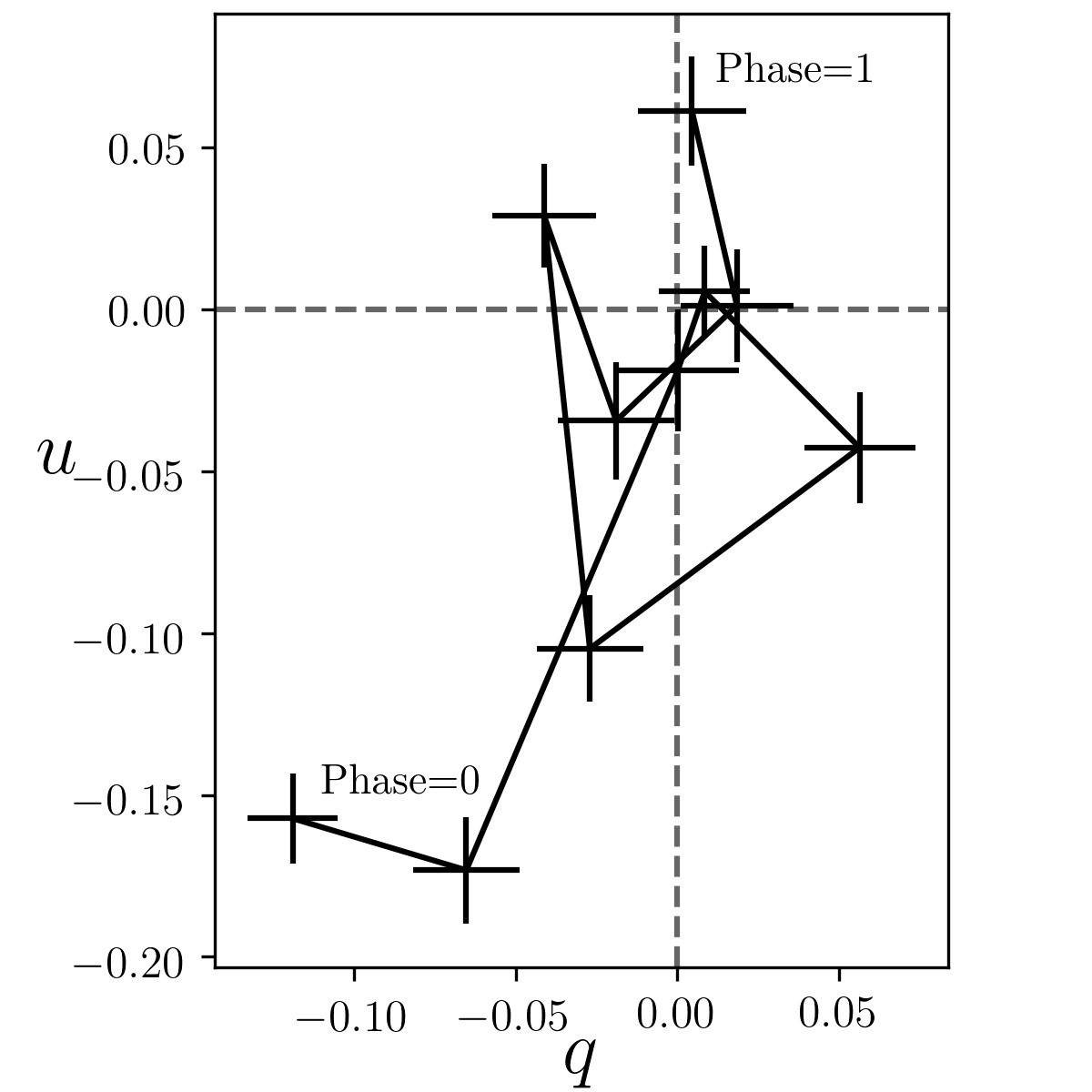}
    \caption{\textit{pfew} Continuum $q$-$u$ observations, phase folded to a 97 day period, and re-binned to 10 phase bins. Solid line shows path between adjacent bins in phase-space.}
    \label{fig:qutrack}
\end{figure}

However, the observations in $q$-$u$ space do not behave like typical binary systems. Figure \ref{fig:qutrack} shows the somewhat erratic path that H$\alpha$ continuum observations take as a function of phase when phase folded by a 97 day period, which shows the most orderly path of all periods in Table \ref{tab:per}. Typically, polarization in binary systems creates elliptical patterns in q-u space \citep{Brown1978PolarisationInclinations}. The path in Fig. \ref{fig:qutrack} is not obviously elliptical; this may be due to the irregular cadence and very long-term data set compared with this short period.

While this does not prove that P Cygni is \textit{not} a binary, it does cast doubt on a simple binary model of the P Cygni system.

\subsection{The Mauchly Test in Polarimetry}
The shape of distributions in $q$-$u$ space encodes important information about the geometry of objects with time varying polarization. For example, N01 discusses how polarimetric observations of Be stars are generally constrained to a line in $q$-$u$ space. The Mauchly Test would find these systems quickly and computationally efficiently.

It is important to note two strengths of the Mauchly test:
\begin{itemize}
    \item This test works even in the case of \textit{unresolved} features.
    \item Since this technique only measures the shape of a distribution, as long as the data being analyzed is at one wavelength, ISP-correction has no effect.
\end{itemize}

Additionally, for objects that are already known to have a preferred position angle, this statistic can quantify how ``preferred'' it really is. In the case of a star with a circumstellar disk, ellipticity of polarimetric observations could yield the inclination angle of the disk with respect to the line of sight. Here, $\mathcal{L}_e = 0$ corresponds to viewing the disk edge on and $\mathcal{L}_e = 1$ corresponds to a face on alignment. 

The eagle-eyed readers may ask, ``\textit{But what about $\beta$ Lyr? That is an eclipsing binary system \citep{Hoffman1998,Lomax2012GEOMETRICALLYRAE}, but your analysis found $\mathcal{L}_e=0.88$. Are you claiming that we are somehow seeing the $\beta$ Lyr system face on?}". Thankfully we are not making that claim. There is some nuance in how we must interpret this statistical tool. $\beta$ Lyr, for which we found $\mathcal{L}_e=0.88$ and $P(\mathcal{L}_e) = 0.0003$, is an excellent example. These values are measurements of our data, \textit{not} of the parent distribution. What these values tell us is that it is very unlikely the the ellipticity of the parent distribution is \textit{greater than} $0.88$. We have placed an upper bound on the true value of $\mathcal{L}_e$.

\section{Conclusion} \label{sec:con}
Using 76 observations taken over 13 years we have been able to update and confirm the conclusions of past work on the circumstellar environment around P Cygni.
\begin{itemize}
    \setlength\itemsep{0em}
    \item P Cygni produces intrinsic polarization, indicating time varying asymmetries in the geometry of the circumstellar material. 
    \item Using the line center method at H$\alpha$ and by fitting the Serkowski law we find a $\lambda_{max}$ of 4595 \AA{} and a $P_{max}$ of 1.23\% for the ISP. This implies that in the direction of P Cygni $R_V = 2.638 \pm 0.028$.
    \item We utilize the Mauchly ellipticity statistic to find that the ellipticity of the distribution of observations in $q$-$u$ space is small and statistically insignificant. Thus we do not see evidence for a preferred position angle.
    \item H$\alpha$, while being less polarized than the continuum on average, contains a wavelength dependent polarization feature. Given its appearance across multiple \textit{HPOL} observations, and in \citet{Davies2005AsphericityVariables}, this feature is most likely astrophysical in nature.
    \item There are a set of statistically significant periods in various aspects of our data. These periodicities are in the $V$-Band, and H$\alpha$ Continuum data, but not in the H$\alpha$ emission line polarization.
\end{itemize}

The feature found in H$\alpha$ is particularly intriguing, for many reasons. It provides a cautionary tale for those wishing to use the line-center method. While we do not believe this structure has had a significant impact on the ISP estimate derived in \S\ref{sec:isp}, it exemplifies how the base assumptions of the line-center method---that strong emission lines in hot stars are intrinsically unpolarized---can break down. This strange feature is also a reminder of how much is left to discover, even about an object like P Cygni which has been studied for more than 400 years, and even about its H$\alpha$ line which has become the prototype for an entire class of similar spectral features. In this sense, this study joins a long and distinguished astronomical tradition, as \citet{1936ApJ....84..296W} wrote after finding variability in multiple absorption features in the spectra of P Cygni taken over the course of 20 years at Lick Observatory, ``...although these variations have hitherto escaped observation, they are by no means minor, but constitute one of the major features of the problem."

More complete and detailed observations of P Cygni will lead to a better understanding of its spectropolarimetric behaviour. The current available data on P Cygni shows its long term behavior. However, it would be incredibly valuable to obtain higher cadence data, particularly, over the course of the longer periods discussed in \S\ref{sec:per}. Such a campaign could reveal potential correlations between the different aspects discussed here and provide a more unified view of the mass-loss of P Cygni. Additionally, observing the ejection of an inhomogeneity, and the $\sim$14 day period afterwards---which \citet{Hayes1985VariableCygni} called the characteristic timescale of polarimetric changes in P Cygni---could reveal much about the nature of the mass-loss in this system.

\acknowledgments
We gratefully acknowledge the contributions of the large team of HPOL and WUPPE observers from the University of Wisconsin who helped to acquire and reduce these data. We also thank Inger Olovsson, curator at the Collections of Skokloster Castle, for her timely and enthusiastic assistance in locating and confirming the existence of Blaeu's globe chronicling the discovery of P Cygni. This research has made use of the SIMBAD database, operated at CDS, Strasbourg, France. Some of the data presented in this paper were obtained from the Mikulski Archive for Space Telescopes (MAST). STScI is operated by the Association of Universities for Research in Astronomy, Inc., under NASA contract no. NAS5-26555. Support for MAST for non-HST data is provided by the NASA Office of Space Science via grant no. NNX09AF08G and by other grants and contracts. This work was funded by Washington NASA Space Grant Consortium, NASA Grant \#NNX15AJ98H and a Cottrell Scholar Award to EML by the Research Corporation for Scientific Advancement.

The authors acknowledge that the work presented was largely conducted on the traditional land of the first people of Seattle, the Duwamish People past and present, and honor with gratitude the land itself and the Duwamish Tribe.

\software{Python 3.7.4, \texttt{Astropy} v4.0.1 \citep{Robitaille2013Astropy:Astronomy,Price-Whelan2018ThePackage}, \texttt{Scipy} v.1.18.1 \citep{Jones2001SciPy:Python}, \texttt{Numpy} v1.17.2 \citep{2011CSE....13b..22V},\texttt{Matplotlib} v3.1.1 \citep{2007CSE.....9...90H}, \texttt{REDUCE} \citep{Nook1990TheObservations,Wolff1996ASightlines,Davidson2014TheObservatory}}

\bibliographystyle{aasjournal}
\bibliography{main}
\end{document}